\documentclass[english, 12pt, a4paper, 
numbers=noenddot
]{scrartcl}
\usepackage{comment} 
\usepackage[T1]{fontenc}
\usepackage[utf8]{inputenc}  
\usepackage[english]{babel} 
\usepackage{lmodern}
\usepackage{amsmath, amssymb, amsthm, amstext} 
\usepackage{thmtools} 
\usepackage{bm}
\usepackage[dvipsnames]{xcolor}
\usepackage{float}
\usepackage[section]{placeins} 
\usepackage{multirow}
\usepackage{booktabs}
\usepackage{longtable}
\usepackage{tabularx} 
\newcolumntype{Y}{>{\centering\arraybackslash}X}
\usepackage{dcolumn}
\newcolumntype{d}[1]{D{.}{.}{#1}}
\usepackage{siunitx}
\usepackage{pgfplotstable}
\usepackage{graphicx}
\usepackage{epstopdf}
\graphicspath{{./figures/}}
\usepackage[babel, english = british]{csquotes} 
\usepackage[backend=biber,
			style=apa, 
			sorting = nyt,
			maxnames = 2,
            maxbibnames=8,
            minbibnames=4,
			doi=false,
			isbn=false,
			url=false,
            dashed=false,
			eprint=false,
            uniquename=false
            ]{biblatex}
\DeclareDelimFormat{nameyeardelim}{\addcomma\space} 
\usepackage[inline]{enumitem}
\usepackage{varioref}
\definecolor{darkblue}{rgb}{0.0, 0.0, 0.55} 
\usepackage[pdfencoding=auto, psdextra, 
   hyperfootnotes=false, 
   colorlinks=true,
   linkcolor=darkblue,   
   citecolor=darkblue,  
   urlcolor=darkblue     
   ]{hyperref} 
\hypersetup{pdfauthor = {Martin C. Arnold and Thilo Reinschluessel},
            pdftitle = {Bootstrap Adaptive Lasso Solution Path Unit Root Tests}
}
\usepackage{nameref}
\usepackage[capitalize, nameinlink]{cleveref} 
\input{ee.sty} 
\usepackage[onehalfspacing]{setspace}
\usepackage{scrlayer-scrpage}
\makeatletter
\newtheoremstyle{Definition}
  {0.2cm}                 
  {0.2cm}                 
  {\normalfont}           
  {}                      
  {\bfseries}  						
  {.}                     
  { }              				
  {}
\newtheoremstyle{Theorem}
  {0.2cm}                 
  {0.2cm}                 
  {\itshape}           	  
  {}                      
  {\bfseries}  						
  {.}                     
  { }              				
  {}
\newtheoremstyle{theoremApostrophe}
  {}                 
  {0.2cm}                 
  {\itshape}           	  
  {}                      
  {\bfseries}  						
  {.}                     
  { }              				
  {\thmname{#1}\thmnumber{ #2}\thmnote{ \normalfont (#3)}}
\theoremstyle{Theorem}
    	\Crefname{cor}{Corollary}{Corollaries}
		\Crefname{prop}{Proposition}{Propositions}
		\Crefname{lem}{Lemma}{Lemmas}
    	\Crefname{thm}{Theorem}{Theorems}
		\Crefname{assum}{Assumption}{Assumptions}
	
		\Crefname{conjecture}{Conjecture}{Conjectures}
		\Crefname{defn}{Definition}{Definitions}
	
		\Crefname{condition}{Condition}{Conditions}

\theoremstyle{Definition}
    	\Crefname{example}{Example}{Examples}	
    \newtheorem{rem}{Remark}
    	\Crefname{rem}{Remark}{Remarks}      
		\Crefname{algo}{Algorithm}{Algorithms}
		
\theoremstyle{Theorem}

\crefname{assumApo}{Assumption}{Assumptions}
\Crefname{assumApo}{Assumption}{Assumptions}

\renewcommand*{\theenumi}{\arabic{enumi}}
\renewcommand*{\theenumii}{(\alph{enumii})}
\renewcommand*{\theenumiii}{(\roman{enumiii})}

\renewcommand*{\p@enumii}{\theenumi\,}
\renewcommand*{\p@enumiii}{\p@enumii.\theenumii} 
\renewcommand*{\p@enumiv}{\p@enumiii.\theenumiii}
\allowdisplaybreaks
\makeatother
\newsavebox\extrainfobox

\newcommand{\DOI}[1]{%
  \gdef\@DOI{DOI: #1}%
}

\addtokomafont{disposition}{\rmfamily}
\addtokomafont{title}{}
\addtokomafont{author}{\large}
\addtokomafont{date}{\large}
\addtokomafont{date}{\vspace*{0.75em}}
\title{Bootstrap Adaptive Lasso Solution Path Unit Root Tests\protect\footnotemark\footnotetext{We thank Christoph Hanck for valuable comments that significantly improved the paper. The authors acknowledge partial financial support from TRR 391 Spatio-temporal Statistics for the Transition of Energy and Transport (520388526) by the Deutsche Forschungsgemeinschaft (DFG, German Research Foundation).}
}
\author{Martin C. Arnold\thanks{Faculty of Business Administration and Economics, University of Duisburg-Essen, Universit\"atsstra{\ss}e 12, 45117 Essen, Germany}\\{\vspace{-1ex}\small \href{mailto:martin.arnold@vwl.uni-due.de}{martin.arnold@vwl.uni-due.de}}  \and Thilo Reinschlüssel\footnotemark[2]~\thanks{RGS Econ - Ruhr Graduate School in Economics, Research Academy Ruhr, Universitätsstra{\ss}e 150, 44801 Bochum, Germany}\\{\small \href{mailto:thilo.reinschluessel@vwl.uni-due.de}{thilo.reinschluessel@vwl.uni-due.de}}}
\date{October 12, 2026} 
\begin{document}
\maketitle
\vspace*{-2em}
\begin{abstract}
    \noindent\textit{\textbf{Abstract}}\\[.5ex]
    We propose sieve wild bootstrap analogues to the adaptive Lasso solution path unit root tests of \textcite{Arnold2024} to improve finite sample properties and extend their applicability to a generalised framework, allowing for non-stationary volatility. Numerical evidence shows the bootstrap to improve the tests' precision for error processes that promote spurious rejections of the unit root null, depending on the detrending procedure. The bootstrap mitigates finite-sample size distortions and restores asymptotically valid inference when the data features time-varying unconditional variance. We apply the bootstrap tests to real residential property prices of the top six Eurozone economies and find evidence of stationarity to be period-specific, supporting the conjecture that exuberance in the housing market characterises the development of Euro-era residential property prices in the recent past.
    \\[2ex]
 \end{abstract}
\renewcommand*{\thefootnote}{\arabic{footnote}}
\setcounter{footnote}{0}
\newpage
\section{Motivation}
In \textcite{Arnold2024}, we propose the $\tau$ and $\Breve{\tau}$ adaptive Lasso unit root tests, which exploit distinct stochastic orders of the activation knots of the lagged level regressor on the Lasso solution path in stationary and non-stationary autoregressions. Simulations reveal a decline in the tests’ precision when adjusting for deterministic components in the presence of higher-order serial correlation. As is well-documented for established unit root tests \parencite[cf.][]{NgPerron2001}, $\tau$ and $\Breve{\tau}$ also are vulnerable to large negative moving average (MA) coefficients, which cause significant upward size distortions. Time-varying variance in the error process can exacerbate such distortions. The latter invalidates the tests' homoskedastic limiting null distributions, further promoting spurious rejections.

A well-established strand of the literature on time series regressions addresses inference on model parameters when the generating process affects a test's distribution with nuisance terms. Several contributions to the unit root literature employ the bootstrap for first-order approximation of the null distribution to improve the finite-sample precision and ensure the asymptotic validity of a test. For autoregressive (AR) models, examples are recursive bootstraps in AR(1) models \parencite{FerrettiRomo1996,Swensen2003}, sieve bootstraps \parencite{ChangPark2003,Park2003} and the residual-based block bootstrap \parencite{PaparoditisPolitis2003}. \textcite{CavaliereTaylor2008,CavaliereTaylor2009heteroskedastic,CavaliereTaylor2009} devise wild bootstrap \parencite{Liu1988} tests for asymptotically valid inference under conditionally heteroskedastic innovations as well as general forms of a time-varying unconditional variance. They propose analogues of the $M$ unit root tests suggested by \textcite{Stock1999,NgPerron2001,PerronNg1996}, the tests of \textcite{Phillips1987} and \textcite{PhillipsPerron1988} and the augmented Dickey-Fuller (ADF) tests of \textcite{SaidDickey1984,Elliottetal1996}.

Bootstraps have also been extensively applied in the recent literature on inference for penalised estimators. \textcite{Chatterjee2011} utilise a residual-based bootstrap to reliably estimate both the distribution and bias of adaptive Lasso estimators in a cross-sectional setting. Similarly, \textcite{Audrino2017} use a residual-based block bootstrap for heteroskedasticity-robust inference in adaptively $\ell_1$-penalised time series regressions. \textcite{Chernozhukov2013} sparked active research by applying the multiplier (wild) bootstrap to penalised estimation in a high-dimensional heteroskedastic setup. Notably, they resample factors that shape the asymptotic distribution, excluding terms that drive asymptotically negligible nuisance parameters by not recalculating the estimator \parencite[the Dantzig selector of ][]{CandesTao2007} during each bootstrap cycle. \textcite{Hansen2018} point out that this procedure is computationally efficient to obtain asymptotically valid inference or confidence intervals at the cost of failing to \enquote{capture any finite sample uncertainty introduced in the
lasso selection}. Contrary to this minimalistic approach, \textcite{Dezeure2017} estimate the distribution of a de-sparsified Lasso estimator for high-dimensional regression by computing the Lasso solutions in every iteration of the wild bootstrap. They assess the computational burden as feasible due to the advances in parallel computing.

In this paper, we address the tests' reliability issues outlined above and propose bootstrap analogues to $\tau$ and $\Breve{\tau}$, using a wild bootstrap scheme for resampling Lasso solution paths in ADF regressions. The method exploits the computational efficiency of the LARS algorithm to compute the Lasso solution path in every bootstrap iteration, similar to the method of \textcite{Dezeure2017}. Our wild bootstrap algorithm follows \textcite{Cavaliereetal2015} and is grounded on the distribution theory, in particular the bootstrap invariance principle, for unconditional heteroskedastic autoregressions of \textcite{CavaliereTaylor2008,CavaliereTaylor2009}. We provide numerical evidence that the wild bootstrap attenuates finite sample size distortions from challenging AR and MA error processes, especially under detrending. Simulations further indicate that---contrary to the unadjusted tests---the wild bootstrap permits valid inference when there is non-stationary volatility in the innovations, mirroring the results of \textcite{CavaliereTaylor2008,CavaliereTaylor2009heteroskedastic} for bootstrap variants of established unit root tests.

The remainder of this article is structured as follows. In \Cref{sec:setup_BALURT}, we discuss the theoretical setup and recap the activation knot unit root tests of \textcite{Arnold2024}. \Cref{sec:wba_BALURT} discusses the wild bootstrap algorithm and its implementation. Monte Carlo studies in \Cref{sec:mce_BALURT} investigate finite sample properties of the bootstrap tests. We illustrate applying the methods using real residential property prices for selected OECD countries in \Cref{sec:ea_BALURT}. \Cref{sec:conclusion_BALURT} concludes and motivates avenues for further research.

We will use the following conventions throughout the manuscript. $\mathbb{I}(\cdot)$ is the indicator function and $\lfloor\cdot\rfloor$ denotes the integer part of its argument. We define $\lVert\vx\rVert_q$ as the $\ell_q$ norm of a vector $\vx$. Coefficients of the true ADF model have a superscript $\star$. A superscript $*$ signifies bootstrap quantities conditional on the observed data. $\xrightarrow[]{d^*}_p$ means bootstrap weak convergence in probability. Convergence in probability and distribution are denoted by $\xrightarrow{p}$ and $\xrightarrow{d}$, respectively.

\section{Setup and Adaptive Activation Knot Unit Root Tests}
\label{sec:setup_BALURT}
\subsection{Setup}
We consider time series $y_t$ generated by an AR data-generating process (DGP) 
	 \begin{align}
        \label{eq:thedgp_BALURT}
        y_t = \vz_t'\vtheta + x_t, \qquad x_t = \varrho x_{t-1} + u_t, \qquad t=0,\dots,T,
    \end{align}
    where $\vz_t := (1,t,\dots,t^m)'$ is an $m^\text{th}$ order deterministic component with coefficient vector $\vtheta$ and $x_t$ is an AR(1) processes with errors $u_t$. The stochastic component $x_t$ satisfies $\varrho\in(-1, 1]$, i.e., $y_t$ is stationary  when $\lvert\varrho\rvert<1$ and has a unit root when $\varrho=1$. We make the following assumptions on the error term $u_t$, cf. Assumption $\mathcal{A}$ in \textcite{CavaliereTaylor2008}.
 
     \begin{restatable}[Linear process errors]{assum}{lperrors_BALURT}
        \label{assum:lperrors_BALURT}
        \noindent
        \begin{enumerate}
        	\item $u_t = \phi(L)\varepsilon_t$ with $\varepsilon_t := \sigma_t \epsilon_t$. The lag polynomial $\phi(L)$ satisfies $\phi(z)\neq0$ for all $\lvert z\rvert\leq1$ and $\sum_{j=1}^\infty j\lvert\phi_j\rvert<\infty$.
        		\label{item:assum:lperrors_BALURT:P1}
        	\item $\epsilon_t$ is a martingale difference sequence (MDS) w.r.t.~the sigma algebra $\mathcal{F}_t := \left\{\epsilon_s, s\leq t\right\}$ such that $\E(\epsilon_t^2) = \sigma_{\epsilon}^2=1$, $T^{-1}\sum_t \epsilon_t^2 \xrightarrow{p} \sigma_{\epsilon}^2$ and $\E\lvert\epsilon_t\rvert^r<K_r$ with $r\geq4$ and some $K_r<\infty$, for all $t$.\label{item:assum:lperrors_BALURT:P2}
        \end{enumerate}
     \end{restatable}
     \begin{restatable}[Unconditional homoskedasticity]{assum}{homoerrors_BALURT}
	     \label{assum:homoerrors_BALURT}
		The unconditional volatility function $\sigma_t$ satisfies $\sigma_t=\sigma\in(0,\infty)\ \,\forall\,t$.	
	 \end{restatable}
\Cref{assum:lperrors_BALURT} comprises a set of standard conditions \parencite[see][Assumption 1]{ChangPark2002}, ensuring that $u_t$ is a weakly-stationary and invertible moving average (MA) process in the $\varepsilon_t$ with finite fourth-order moments. Together with \Cref{assum:homoerrors_BALURT}, which requires the error process $u_t$ to be unconditionally homoskedastic, \Cref{assum:lperrors_BALURT} imposes the same conditions on $u_t$ as we do in Assumption 1 of \textcite{Arnold2024}. We note that the MDS condition in part~\ref{item:assum:lperrors_BALURT:P2} allows for conditionally heteroskedastic $\varepsilon_t$, e.g., weakly stationary generalised autoregressive conditionally heteroskedastic (GARCH) and Markov-switching processes, but excludes unconditionally heteroskedastic error processes. The latter is allowed for by the following generalisation of the volatility function in \Cref{assum:homoerrors_BALURT}.
	 	\begin{restatable}[Unconditional heteroskedasticity]{assumApo}{heteroerrors_BALURT}
		 	\label{assum:heteroerrors_BALURT}
			The unconditional volatility function $\sigma_t$ satisfies $\sigma_{\lfloor rT\rfloor}=\omega(r),\, r\in[0,1]$ with $\omega(r)$ a non-stochastic càdlàg function such that $0<\sigma_t<\infty\ \forall\,t$.	
		\end{restatable}
		\Cref{assum:heteroerrors_BALURT} is a relaxation of \Cref{assum:homoerrors_BALURT} and allows $\sigma_t$ to exhibit non-stochastic time-varying volatility of a quite general form, including (a countable number of) abrupt jumps, polynomial trends and smooth transitions in the unconditional variance. It also generalises the setup of \textcite{Arnold2024} towards a broader class of empirical processes relevant to macroeconomic applications. We refer to Remarks 1 and 2 in \textcite{CavaliereTaylor2008} and Section 2 in \textcite{CavaliereTaylor2009} for a discussion of heteroskedastic error processes covered by allowing for MDS innovations $\epsilon_t$ as in part \ref{item:assum:lperrors_BALURT:P2} of \Cref{assum:lperrors_BALURT} and deterministic non-stationary volatility implied by \Cref{assum:heteroerrors_BALURT}.

Next, we will review the adaptive Lasso activation knot tests of \textcite{Arnold2024}.

\subsection{Activation Knot Unit Root Tests}
Based on the ADF($\infty$) representation of \eqref{eq:thedgp_BALURT} if $\vz_t'\vtheta = 0$,
\begin{align}
    \Delta y_t =& \, \rho^\star y_{t-1} + \sum_{j=1}^{\infty} \delta_j^\star \Delta y_{t-j} + \varepsilon_{t},\label{eq:adfmod_BALURT}
\end{align} 
with $ \rho^\star\in(-2,0]$, $\sum_{j=1}^\infty \delta_j^\star < 1$, we consider the approximating ADF($p$) regression model
\begin{align}
    \Delta y_t =& \, \rho_{p} y_{t-1} + \sum_{j=1}^p \delta_{p,\,j} \Delta y_{t-j} + \varepsilon_{p,\,t}.\label{eq:adfreg_BALURT}
\end{align}
The AR lag order $p$ in the model \eqref{eq:adfreg_BALURT} meets the following assumption.
	\begin{restatable}[Lag order]{assum}{aroder_BALURT}
		\label{assum:aroder_BALURT}
		$p$ satisfies $p=o(T^{1/3})$ and $p\to\infty$ as $T\to\infty$.
	\end{restatable}
	In \textcite{Arnold2024}, we investigate testing for a unit root using the adaptive Lasso solution path to \eqref{eq:adfreg_BALURT}. Consider the adaptively penalised loss function
	\begin{align}
		\Psi_T(\dot\rho,\dot\vdelta\vert\lambda) := \sum_t \left(\Delta y_t - \dot\rho y_{t-1} - \sum_{j=1}^p \dot\delta_j \Delta y_{t-j} \right)^2 +&\, 2\lambda\left( w_1^{\gamma_1} \lvert\dot\rho\rvert + \sum_{j=1}^p w_{2,\,j}^{\gamma_2} \left\lvert\dot\delta_j\right\rvert\right),
    \label{eq:adaptive_lassooptim_BALURT}
  \end{align}
  with adaptive weights $w_1 := 1/\lvert\widehat{\rho}\rvert$, $w_{2,\,j} := 1/\left\lvert\widehat{\delta}_j\right\rvert$ determined by the initial OLS estimates $\widehat{\rho}$ and $\widehat{\delta}_j$ in \eqref{eq:adfreg_BALURT}, and adjustment parameters $\gamma_1,\gamma_2\in\mathbb{R}^+$. The adaptive Lasso solution path to \eqref{eq:adaptive_lassooptim_BALURT},
	\begin{align}
		\mathcal{L} := \left\{(\widehat{\rho}_\lambda, \widehat\vdelta_\lambda')'\ \bigg\vert \
		(\widehat{\rho}_\lambda, \widehat\vdelta_\lambda')' := \arg\min_{\dot\rho,\, \dot\vdelta} \Psi_T(\dot\rho,\dot\vdelta\vert\lambda), \ \ \lambda\in\mathbb{R}^+ \right\}, \label{eq:ALestimator_path_BALURT}
	\end{align}
 	is a collection of solutions $\widehat{\vbeta}_\lambda := (\widehat{\rho}_\lambda, \widehat\vdelta_\lambda')'$ subject to the $\ell_1$ penalty parameter $\lambda$. A solution path $\mathcal{L}$ is characterised by knots at which variables are activated or deactivated. Our testing principle leverages that a unit root causes the activation knots of $y_{t-1}$ to be of different stochastic order than under stationarity. We propose a right-sided test of $H_0:\rho^\star=0$ against $H_1:\rho^\star\in(-2,0)$ via the statistic $\tau_{\gamma_1} := T^{\gamma_1-1} \lambda_{0,\,\rho^\star}$ with $\lambda_{0,\,\rho^\star}$, the first activation knot of $y_{t-1}$, satisfying
    \begin{align*}
        \lambda_{0,\,\rho^\star} = w_1^{-\gamma_1} \left\lvert \sum_t y_{t-1} \left(\Delta y_t - \sum_{j=1}^{p}  \widehat{\delta}_{\lambda,\,j} \Delta y_{t-j} \right)\right\rvert.
    \end{align*}
  The knot $\lambda_{0,\,\rho^\star}$ is a standard output of algorithms for calculating \eqref{eq:ALestimator_path_BALURT} such as LARS \parencite{Efronetal2004}, making $\tau_{\gamma_1}$ straightforward to compute.\footnote{We refer to Section 3.1 in \textcite{Arnold2024} for the formal definition of an activation knot and a discussion of the stochastic properties of $\lambda_{0,\,\rho^\star}$ in particular.} 
 	
 	Given consistency of the estimator $\widehat{\vbeta}_\lambda$ at $\lambda = \lambda_{0,\,\rho^\star\sim c/T}$, the activation knot of $y_{t-1}$ under local-to-unity roots with non-centrality parameter $c\in(-\infty,0]$, the (local) distribution of $\tau_{\gamma_1}$ can be identified. Specifically, for $\varrho^\star=1+c/T$ and $\gamma_1>1/2,\ \gamma_2>0$, Theorem 1 in \textcite{Arnold2024} states that for unconditionally homoskedastic $u_t$ satisfying \Cref{assum:lperrors_BALURT},
	\begin{align}
    	\tau_{\gamma_1} \xrightarrow{d} \mathcal{T}_{c,\,\gamma_1} := \left\lvert \phi(1)^{-1} \frac{W_c(1)^2-1}{2\int_0^1 W_c(r)^2\mathrm{d}r}\right\rvert^{\gamma_1} \biggl\lvert\frac{1}{2}\sigma^2\phi(1)\left(W_c(1)^2-1\right)\biggr\rvert,\label{eq:taugammaNURLimit_BALURT}
    \end{align}
	where $W_c(r) := \int_0^r \mathrm{exp}[c(r-s)]\mathrm{d}W(s)$ is an Ornstein-Uhlenbeck process driven by the standard Wiener process $W(s)$ on $s\in[0,1]$.
	
	A modification of $\tau_{\gamma_1}$ proposed in \textcite{Arnold2024} derives from enhancing the penalty weight for $y_{t-1}$ with additional information on whether $\rho^\star=0$. This information enrichment of $w_1$ proceeds as $\Breve{w}_1 := w_1 \cdot J_\alpha$, using the statistic $J_\alpha$ \parencite[see][]{HerwartzSiedenburg2010} which exploits different stochastic orders of the OLS estimator in time series regressions when the degree of integration differs. Analogous to $\tau_{\gamma_1}$, the modified statistic is calculated as $\Breve{\tau}_{\gamma_1} := T^{\gamma_1-1}\Breve{\lambda}_{0,\,\rho^\star}$, where $\Breve{\lambda}_{0,\,\rho^\star}$ denotes the first activation knot of $y_{t-1}$ on a solution path $\Breve{\mathcal{L}}$ based on $\Breve{w}_1$.\footnote{See Algorithm 1 in \textcite{Arnold2024} for details on the computation of $J_\alpha$.} By Corollary 1 of \textcite{Arnold2024}, 
	\begin{align}
    	\Breve{\tau}_{\gamma_1} \xrightarrow{d} \Breve{\mathcal{T}}_{c,\,\gamma_1} := \mathcal{T}_{c,\,\gamma_1}/J_{\alpha,\,c}^{\gamma_1},\label{eq:brevetaugammaNURLimit_BALURT}
    \end{align}
	where $J_{\alpha,\,c}$ is the $c$-dependent limit of $J_\alpha$.

	For implementation we propose the natural choice $\gamma_1=\gamma_2=1$ which avoids an adjustment for $\phi(1)\neq1$ and yields tests with limit distributions
	\begin{align}
		\begin{split}
			\tau := \tau_{\gamma_1=1} / \widehat{\sigma}^2 \ \xrightarrow{d}&\,\ \frac{\left(W_c(1)^2-1\right)^2}{4\int_0^1 W_c(r)^2\mathrm{d}r},\\ 
			\Breve{\tau} := \Breve{\tau}_{\gamma_1=1} / \widehat{\sigma}^2 \ \xrightarrow{d}&\,\ \frac{1}{J_{\alpha,\,c}}\frac{\left(W_c(1)^2-1\right)^2}{4\int_0^1 W_c(r)^2\mathrm{d}r},
		\end{split}\label{eq:gamma1limit_BALURT}
	\end{align}
	where $\widehat\sigma^2$ estimates the error variance based on OLS residuals from the penalty weights ADF regression \eqref{eq:adfreg_BALURT}.
	
	To accommodate for $\vz_t\neq\vzero$, standard detrending ideas can be applied before computing the Lasso solution. In the remainder, we follow \textcite{Arnold2024} and consider first-difference (FD) detrending \parencite{SchmidtPhillips1992} before computing the Lasso solutions and calculating $J_{\alpha}$ on OLS-adjusted data. Under detrending, the $W_c(r)$ in \eqref{eq:gamma1limit_BALURT} are replaced by the corresponding projection, resulting in limit distributions deviating from the case without adjustment for deterministic components. The $\Breve{\tau}$ limit is further affected by the distribution of $J_{\alpha,\,c}$, which is OLS adjusted. Critical values for $\tau$ and $\breve{\tau}$ under FD adjustment for deterministic components are reported in Table D1 in \textcite{Arnold2024}. 
	
    In the next section, we summarise the implications of \Cref{assum:heteroerrors_BALURT} and discuss a wild bootstrap correction that addresses the ramifications for the activation knot tests.

\section{Wild Bootstrap Activation Knot Tests}
\label{sec:wba_BALURT}

While \textcite{Arnold2024} find that $\tau$ and $\Breve{\tau}$ have mostly good size, simulation studies indicate some downward or upward size distortions in small samples for AR error processes with high autocorrelation or processes with MA roots close to $-1$. These distortions are somewhat more pronounced under detrending. 

As we demonstrate in \Cref{sec:mce_BALURT}, heteroskedastic error processes may amplify these undesirable finite sample properties. Unconditional heteroskedasticity invalidates inference using the homoskedastic null distributions in \eqref{eq:gamma1limit_BALURT}. \textcite{CavaliereTaylor2007} show that non-stationary volatility as permitted under \Cref{assum:heteroerrors_BALURT} alters the limit distributions of common (unpenalised) regression-based unit root tests in \eqref{eq:adfreg_BALURT} as non-stationary volatility introduces nuisance parameters to the tests' homoskedastic distributions in \eqref{eq:gamma1limit_BALURT} that do not vanish asymptotically, invalidating the critical values based on the limits for $c=0$. For DGP \eqref{eq:thedgp_BALURT} in the local-to-unity case $\varrho^\star = 1+c/T$ with $-\infty<c\leq0$, this can be traced back to the invariance principle
	\begin{align}
		T^{-1/2} y_{\lfloor rT\rfloor} \ \xrightarrow{d} \ \overline{\omega}\phi(1)W_{c,\,\eta}(r), \quad r\in[0,1], \label{eq:nsvFCLT_BALURT}
	\end{align}
	depending on the volatility function $\omega(\cdot)$, cf. the discussion of Theorem 1 in \textcite{CavaliereTaylor2007}. Here, $\overline{\omega}^2 := \int_0^1 \omega(s)^2\mathrm{d}s$ is the limit of $T^{-1}\sum_{t=1}^T\sigma_t^2$ and 
	\begin{align}
		W_{c,\,\eta}(r) := \int_0^r \mathrm{exp}[c(r-s)]\mathrm{d}W_{\eta}(s),\quad r\in[0,1],
	\end{align}
	is a diffusion process driven by a time-transformed Wiener process $W_{\eta}(\cdot) := W(\eta(\cdot))$ with directing process $\eta(\cdot)$. A key quantity is the functional 
	\begin{align}
		\eta(r) := \overline{\omega}^{\,-2}\int_0^r\omega(s)^2\mathrm{d}s, \quad r\in[0,1],
	\end{align}
	the so-called \emph{variance profile} of the series, cf. Section 3 in \textcite{CavaliereTaylor2007}. A non-constant unconditional volatility function $\omega(\cdot)$ thus alters the tests' (asymptotic) null distributions and local power functions, invalidating the local asymptotic results for $\tau$ and $\Breve{\tau}$ in \eqref{eq:gamma1limit_BALURT}. Under constant volatility, $\overline{\omega}^2=\sigma^2\in(0,\infty)$. Therefore, $\eta(s)=s,\, s\in[0,1]$ so that $W_{\eta}(\cdot) = W(\cdot)$ and the limiting r.v.~in \eqref{eq:nsvFCLT_BALURT} reduces to the same (scaled) Ornstein-Uhlenbeck process underlying the limiting functionals in \eqref{eq:gamma1limit_BALURT}.

The wild bootstrap proposed in \textcite{CavaliereTaylor2008} samples bootstrap innovations as $\varepsilon_t^* := \xi_t \cdot \check{\varepsilon}_{p,\,t}^{d}$ with $\check{\varepsilon}_{p,\,t}^{d}$ the OLS residuals from the ADF regression \eqref{eq:adfreg_BALURT} based on the detrended data and the $\xi_t$ are i.i.d.~with $\E(\xi_t) = 0$ and $\Var(\xi_t) = 1$. The resampled data $y_t^*$ generated by the partial sum process
\begin{align*}
    y_t^* = y_0^* + u_t^*, \qquad u_t^* = \sum_{i=1}^t \varepsilon_i^*, \qquad t=1,\dots,T,
\end{align*}
initialised at $y_0^*=0$, then satisfy the bootstrap invariance principle
\begin{align}
    T^{-1/2} y_{\lfloor rT\rfloor}^* \xrightarrow[]{d^*}_p \overline{\omega} W_{0,\,\eta}(r), \quad r\in[0,1],\label{eq:BALURT_binvarp}
\end{align}
cf.~Eq.~(4) in the proof of Theorem 2 of \textcite{CavaliereTaylor2008}. 

Since the $\xi_t$ are i.i.d., the device $\varepsilon^*_t = \xi_t \cdot\check{\varepsilon}_{q,\,t}$ anihilates any serial correlation from the original shocks. While this does not impact the asymptotic validity of the bootstrap tests, as follows from the exposition in \textcite{CavaliereTaylor2008,CavaliereTaylor2009}, neglecting correlation in the original shocks may reduce the finite-$T$ precision of the bootstrap tests. We thus follow \textcite{CavaliereTaylor2009,SmeekesTaylor2012} and build higher-order stationary dynamics estimated from the data into the bootstrap errors,
\begin{align}
    u_t^* = \sum_{j=1}^q \check{\delta}_{q,\,j}\, u_{t-j}^* + \varepsilon^*_t, \quad t=1,\dots,T, \label{eq:recolouring}
\end{align}
and construct the bootstrap sample as
\begin{align*}
    y_t^* := y_0^* + \sum_{i=1}^t u_i^*, \quad 0,\dots,T.
\end{align*}
The $\check\delta_{q,\,j}$, $j=1,\dots,q$ are estimated coefficients of the $\Delta y_{t-j}$ in a sieve ADF regression \eqref{eq:adfreg_BALURT} with lag order $q$, similarly as in the residual-based bootstrap schemes of \textcite{FerrettiRomo1996}, \textcite{ChangPark2003, Park2003}. Setting $q = p$, the recolouring recursion \eqref{eq:recolouring} ensures that\footnote{Other choices for $q$ are discussed in \Cref{rem:bootlagorder}. A proof of \eqref{eq:BALURT_binvarprecol} under Assumptions equivalent to ours is given in the proof of Theorem 2 of \textcite{SmeekesTaylor2012}.}
\begin{align}
    T^{-1/2} y_{\lfloor rT\rfloor}^* = T^{-1/2} \sum_{t=1}^{\lfloor rT\rfloor} u_t^* \xrightarrow[]{d^*}_p \overline{\omega}\phi(1)W_{0,\,\eta}(r), \quad r\in[0,1].\label{eq:BALURT_binvarprecol}
\end{align}
\cref{eq:BALURT_binvarprecol} states that the bootstrap correctly replicates effects from stationary serial correlation, cf. \eqref{eq:nsvFCLT_BALURT}, and thus may better reproduce nuisance terms from adjusting for $\phi(1)\neq 1$ in the finite-$T$ null distributions of $\tau$ and $\Breve{\tau}$. The invariance principles \eqref{eq:BALURT_binvarp} and \eqref{eq:BALURT_binvarprecol}, and continuous mapping arguments are the theoretical foundation for applying the (sieve) wild bootstrap to the activation knot tests $\tau$ and $\Breve{\tau}$ for improving finite sample precision and valid asymptotic inference in the present heteroskedastic framework.

We next present the wild bootstrap algorithm, which adapts the wild bootstrap algorithms for the popular ADF and $M$ tests proposed by \textcite{CavaliereTaylor2008,CavaliereTaylor2009,Cavaliereetal2015} to our activation knot tests.

\begin{restatable}[Wild bootstrap activation knot test]{algo}{abde}
 	\noindent
 	\label{algo:wbftau}
 	\begin{enumerate}
 		\item Adjust $\{y_t\}_{t=0}^T$ for the determinisitic component $\vz_t'\vtheta$ using FD detrending. Denote the resulting series $\{y^{d}_t\}_{t=0}^T$.
 		\item Select a lag order $q$ (see \Cref{rem:bootlagorder}) and obtain the residuals $\{\check{\varepsilon}_{q,\,t}^d\}_{t=1}^T$, where
 		\begin{align}
 			\check{\varepsilon}_{q,\,t}^d := \Delta y_{t}^d - \check{\rho}_{q} y_{t-1}^d - \sum_{j=1}^q \check{\delta}_{q,\,j} \Delta y_{t-j},
 			\label{eq:consistentresiduals}
 		\end{align}
 		using the estimates $\check{\rho}_q,\check{\delta}_{q,\,1},\dots,\check{\delta}_{q,\,q}$, defining $ (y_{-1}^d,\dots, y_{-q}^d)' := \vzero$. Calculate the adaptive Lasso solution path for the model underlying \eqref{eq:consistentresiduals} up to $\lambda = \lambda_{0,\,\rho^\star}$, the \emph{first} activation knot of $y_{t-1}$, and compute $\tau_{\gamma_1}$.
 	\item Generate wild bootstrap innovations $\left\{\varepsilon^*_t\right\}_{t=1}^T$ according to the device $\varepsilon^*_t := \xi_t \cdot\check{\varepsilon}_{q,\,t}^d$, where the r.v.s $\xi_t$ are i.i.d.~and satisfy $\E(\xi_t)=0$ and $\Var(\xi_t)=1$.
 	\item Build the bootstrap error process $\{u_t^*\}_{t=1}^T$ using the recolouring recursion
 		\begin{align}
	 	 u_t^* = \sum_{j=1}^q \check{\delta}_{q,\,j}\, u_{t-j}^* + \varepsilon^*_t,
 		\end{align} 
 		with $(u^*_0,\dots,u^*_{1-q})':=\vzero$. Build a bootstrap time series $\{y_t^*\}_{t=0}^T$ via the partial sum process
 		\begin{align*}
 		 y_t^* := y_0^* + \sum_{i=1}^t u_i^*,
 		\end{align*}
 		with $y_0^* = 0$.
 	\item Adjust the bootstrap sample $\{y_t^*\}_{t=0}^T$ as in Step 1 and compute the bootstrap adaptive Lasso solution path for an ADF regression with lag order $p^*$ (cf. \Cref{rem:bootlagorder}) up to $\lambda = \lambda_{0,\,\rho^\star}^*$, with
    \begin{align*}
        \lambda_{0,\,\rho^\star}^* = \left(\frac{1}{w_1^{*}} \right)^{\gamma_1} \left\lvert \sum_t y_{t-1}^{*d} \left(\Delta y_t^{*d} - \sum_{j=1}^{p^*}  \widehat{\delta}_{\lambda,\,j}^{*} \Delta y_{t-j}^{*d} \right)\right\rvert,
    \end{align*}
    the \emph{first} activation knot of $y_{t-1}^{*d}$, and compute the bootstrap test statistic $\tau^*_{\gamma_1,\,b}$.
 	      \item Obtain bootstrap test statistics $\{\tau^*_{\gamma_1,\,b}\}_{b=1}^B$ by completing steps 3 to 5 $B$ times. Calculate the bootstrap level-$\alpha$ critical value as
                \begin{align}
 		     \textup{CV}^*(\alpha) := \max \left\{x:\ \frac{1}{B}\sum_{b=1}^B \mathbb{I}\left(x<\tau^*_{\gamma_1,\,b}\right)\leq\alpha\right\}.
 		    \end{align}
 		    Reject the unit root null at level $\alpha$ if $\textup{CV}^*(\alpha) \leq\tau_{\gamma_1}$ with $\tau_{\gamma_1}$ the test statistic computed for lag order $p$ using the detrended data of Step 1.\qed
 	\end{enumerate}
 \end{restatable}

\begin{rem}
    \label{rem:bootlagorder}
    The lag orders $p$, $q$, and $p^*$ must be selected to implement \Cref{algo:wbftau}. Modified information criteria such as the MAIC \parencite{NgPerron2001} are established procedures for this purpose. The RSMAIC of \textcite{Cavaliereetal2015} is a heteroscedasticity-robust variant of the MAIC, which estimates the lag order based on a rescaling of the $y_t$ with a non-parametric estimate of its (assumed sufficiently smooth) variance profile, as suggested by \textcite{Beare2018}. Since the power advantage of using the RSMAIC over the MAIC for sieve wild bootstrap ADF tests under non-stationary volatility is well documented by the simulation results in \textcite{Cavaliereetal2015}, with negligible effects under homoskedasticity, we apply the RSMAIC throughout.
    
    \noindent Although $q$ (and $p^*$) are not required to diverge for the asymptotic validity of the bootstrap \parencite[cf.][]{Cavaliereetal2015}, selecting $p$ and $p^*$ by the RSMAIC and setting $q=p$ is a convenient choice. Notably, selecting $p^*$ independently from $p$ and $q$ has been documented to help control upward size distortions of sieve bootstrap ADF tests under errors with a large negative MA(1) component, where information criteria yield underspecified models \parencite{Richard2009}.
\end{rem}

\begin{rem}
     \label{rem:ALresiduals}
     Obtaining the residuals $\check{\varepsilon}_{q,\,t}$ from the sieve regression underlying \eqref{eq:consistentresiduals} using OLS is convenient since we require estimates of $(\rho_p$, $\vdelta_p')'$ for computing the adaptive penalty weights of the Lasso estimator anyhow. \noindent\textcite{CavaliereTaylor2009} suggest other asymptotically equivalent strategies to compute the residuals, which we do not consider here for brevity.
     
     \noindent If OLS estimation is infeasible (e.g., due to collinearity or $p\geq T$) but zero-consistent initial coefficient estimates\footnote{Zero-consistency ensures the penalty weights to be bounded for relevant variables and to converge to infinity for irrelevant variables. Zero-consistency is the weakest requirement for establishing the oracle property in the literature on the adaptive Lasso to our knowledge.} \parencite{Huangetal2008} are available, ad-hoc estimates for $\check{\varepsilon}_t$ can be obtained by running AL or ALIE and setting $\lambda = \lambda^\tau_\alpha$ with $\lambda^\tau_\alpha$ a (upper-tail) quantile of the null distribution of $\tau$. This approach resembles that of \textcite{Chernozhukov2023} and yields consistent coefficient estimates from conservative model selection, as for AIC-tuned estimation. However, finite-$T$ adaptive Lasso estimates are usually biased and require recentering of the residuals \parencite[cf.][]{Chatterjee2011}. Therefore, we reckon using OLS residuals is more convenient for practitioners if feasible.
     
\end{rem}

\begin{rem}
     \label{rem:bdevicedist}
     The literature on the wild bootstrap \parencite[cf.][]{Liu1988,DavidsonFlachaire2008} features several proposals on how to sample the $\xi_t$ in generating the bootstrap errors $\varepsilon_t^*$ in step 3. An example is the asymmetric two-point distribution by \textcite{Mammen1993}, designed for higher precision of bootstrap tests under heteroskedastic and non-Gaussian errors. In other applications, e.g., the pooled panel unit root tests of \textcite{HerwartzWalle2018}, the Rademacher distribution \parencite{DavidsonFlachaire2008} has been reported to yield better precision. However, and consistent with the results for various wild bootstrap time series unit root tests reported in \textcite{CavaliereTaylor2008,CavaliereTaylor2009heteroskedastic,DemetrescuHanck2016}, we find Gaussian $\xi_t$ to yield good performance, with no significant discrepancies to using other candidate distributions. We therefore report results only for Gaussian $\xi_t$. Simulation results for resampling with the Rademacher and Mammen's distribution are available on request.
 \end{rem}

\section{Monte Carlo Evidence}
\label{sec:mce_BALURT}
To investigate sample properties of the tests, we generate time series as  
\begin{equation}
    y_t = \varrho y_{t-1} + v_t,\quad t=0,1,\dots,T, \label{eq:ardgp_BALURT}
\end{equation} 
with starting value $y_0=0$ for sample sizes $T\in\{75,100,150,250,500,1000\}$. We let $\varrho = 1+c/T$ so that setting $c=0$ obtains data under the unit root null. As (local) stationary alternatives, we set $c=-7$ when testing based on data adjusted for a constant and $c=-13.5$ for detrending. Pure AR or MA errors are generated with the recursion 
\begin{align}
	 v_t = \varphi v_{t-1} + \vartheta \epsilon_{t-1} + \sigma_{T,\,t}\cdot\epsilon_t, \quad\epsilon_t\overset{i.i.d.}{\sim}N(0,1),\label{eq:simerrors_BALURT}
\end{align}
with coefficients $\varphi,\vartheta\in\{-.8, -.4,\allowbreak 0, .4, .8\}$. 

Following \textcite{CavaliereTaylor2009heteroskedastic}, we model deterministic smooth transitions in the unconditional volatility parameter $\sigma_{T,\,t}$ between two regimes with variances $s_1^2,s_2^2>0$ via a logistic function $\mathbb{S}_{T,\,t}$,
	\begin{align}
		\sigma^2_{T,\,t} := s_1^2 + (s_2^2 - s_1^2) \cdot \mathbb{S}_{T,\,t}, \quad \mathbb{S}_{T,\,t} := (1 + \exp(-\tilde{\gamma}_T(t-\lfloor \kappa T\rfloor)))^{-1},\label{eq:svt_BALURT}
	\end{align}
	with transition midpoint $\lfloor \kappa T\rfloor$,  $\kappa\in(0,1)$. The parameter $\tilde{\gamma}_T$ determines the transition speed between $s_1^2$ and $s_2^2$, yielding and abrupt regime switch at $t=\lfloor \kappa T\rfloor$ in the sense that $\mathbb{S}_{T,\,t}\to\mathbb{I}(t\geq \lfloor\kappa T\rfloor)$ as $\tilde{\gamma}_T\to\infty$. We adapt the local drift $\tilde{\gamma}_T := 25/T$ from \textcite{CavaliereTaylor2009heteroskedastic}. Setting $s_1^2 = 1$ throughout, we model negative shifts with early transition midpoints ($\kappa = .2$, $s_2^2 = .25$) as well as positive shifts with late midpoints ($\kappa = .8$, $s_2^2 = 4$).
	
The bootstrap variants of $\tau$ and $\Breve{\tau}$, denoted $\tau^*$ and $\Breve{\tau}^*$, are implemented as detailed in \Cref{sec:wba_BALURT} at the 5\% level and computed with $B=499$ bootstrap replications. The $\xi_t$ in Step 3 of \Cref{algo:wbftau} are standard Gaussian r.v.s, and the lag orders $p,q$ and $p^*$ are estimated using the RSMAIC, if not indicated otherwise. We compute the Lasso solution paths using the implementation of the LARS algorithm \parencite{Efronetal2004} in the \texttt{R} (\citeyear{R})   package \texttt{lars} \parencite{pkg-lars}.

We first examine the ability of the wild bootstrap to approximate the tests' finite sample distributions and assess the bootstrap tests' precision and local power in the baseline scenario with uncorrelated homoskedastic errors ($s_2^2=1$). The sieve regression lag order $q$ is zero in this experiment. The results presented in \Cref{tab:simplear_BALURT} indicate that $\tau^*$ and $\Breve{\tau}^*$ have excellent precision (top panel), improving on the empirical size of $\tau$ and $\Breve{\tau}$, which may be somewhat conservative for small $T$. Furthermore, the bootstrap tests seem to replicate the (size-adjusted) local power function of the standard tests quite well, with only minor deviations for small $T$ notable for $\Breve{\tau}$ (bottom panel).

\begin{table}[t]
\setlength{\tabcolsep}{14pt}
\renewcommand{\arraystretch}{1}
\centering
\caption{Rejection rates of the WB activation knot tests for i.i.d.~errors}
\label{tab:simplear_BALURT}
\vspace{.3cm}
\resizebox{\textwidth}{!}{
	
\begin{tabular}{l *{8}{S} }
\toprule
\toprule
\multicolumn{1}{c}{} & \multicolumn{4}{c}{$\vz_t=1$} & \multicolumn{4}{c}{$\vz_t=(1,t)'$} \\
\cmidrule(l{3pt}r{3pt}){2-5} \cmidrule(l{3pt}r{3pt}){6-9}
$T$ & $\tau^*$ & $\Breve{\tau}^*$ & $\tau$ & $\Breve{\tau}$ & $\tau^*$ & $\Breve{\tau}^*$ & $\tau$ & $\Breve{\tau}$\\
\midrule
\addlinespace[0.3em]
\multicolumn{9}{c}{\textbf{$\varrho^\star = 0$}}\\
75 & 0.0488 & 0.046 & 0.0374 & 0.0364 & 0.047 & 0.0468 & 0.0398 & 0.035\\
100 & 0.055 & 0.056 & 0.045 & 0.0436 & 0.0474 & 0.0472 & 0.0406 & 0.0368\\
150 & 0.0506 & 0.0546 & 0.043 & 0.043 & 0.0494 & 0.0482 & 0.0412 & 0.0378\\
250 & 0.0554 & 0.0518 & 0.0488 & 0.0434 & 0.0452 & 0.0466 & 0.0414 & 0.0398\\
500 & 0.0556 & 0.0542 & 0.0504 & 0.0488 & 0.052 & 0.0528 & 0.047 & 0.045\\
1000 & 0.0546 & 0.0496 & 0.0492 & 0.0456 & 0.0522 & 0.0492 & 0.0474 & 0.0448\\
\addlinespace[0.3em]
\multicolumn{9}{c}{\textbf{$\varrho^\star = -c/T$}}\\
75 & 0.2466 & 0.297 & 0.2508 & 0.3116 & 0.3682 & 0.3954 & 0.384 & 0.4112\\
100 & 0.2412 & 0.2882 & 0.2284 & 0.2776 & 0.3654 & 0.3822 & 0.3658 & 0.4032\\
150 & 0.2358 & 0.2966 & 0.2338 & 0.28 & 0.3656 & 0.3816 & 0.374 & 0.404\\
250 & 0.2458 & 0.3078 & 0.227 & 0.304 & 0.3678 & 0.3874 & 0.3894 & 0.4016\\
500 & 0.2512 & 0.307 & 0.232 & 0.2858 & 0.3818 & 0.4026 & 0.3702 & 0.3896\\
1000 & 0.3098 & 0.3594 & 0.296 & 0.351 & 0.3914 & 0.4158 & 0.3816 & 0.418\\
\bottomrule
\end{tabular}
}
\begin{minipage}{\textwidth}
    \vspace{.25cm}
    \scriptsize\textit{Notes:} DGP \eqref{eq:ardgp_BALURT} with $\sigma_{T,\,t}=1\,\forall\,t$. $B=499$ wild bootstrap replications with $q = 0$ (no recolouring). The data are adjusted for a constant or a linear time trend using the FD method of \textcite{SchmidtPhillips1992}. The lag orders $p$ and $p^*$ are selected using the RSMAIC. Top panel: unit root model. 
    Bottom panel: local alternative with $c=-7$ if $\vz_t = 1$ and $c = -13.5$ if $\vz_t = (1,t)'$. Power estimates of $\tau$ and $\Breve{\tau}$ are size-adjusted at 5\%. $5000$ Monte Carlo replications.
    \end{minipage}
\end{table}

In a second experiment, we examine the precision of the sieve bootstrap tests under homoskedastic correlated errors. \Cref{tab:correrr_BALURT} presents the results. The bootstrap tests perform well for AR errors with $\varphi\in(-.8, .8)$ and under MA errors with $\vartheta=.8$. The most challenging scenario is MA errors with $\vartheta = -.8$, which lead to upward size distortions across $T$ that are prohibitive for small samples and only decay slowly with the sample size. Although $\tau^*$ and $\Breve{\tau}^*$ are oversized, resampling with recolouring yields significant improvements over $\tau$ and $\Breve{\tau}$. Recolouring is particularly helpful under detrending, where the empirical size of the bootstrap tests converges quickly towards the $5\%$ level, undercutting the rejection rates of the unadjusted tests by over a third in larger samples. Furthermore, comparing the size of the SWB tests with the size of the WB tests with $q=0$ reported in \Cref{tab:tab_TA5_BALURT,tab:tab_TA6_BALURT} of \Cref{sec:asr_BALURT} corroborates better precision through the sieve step under MA errors in particular.

\begin{table}[htbp]
\setlength{\tabcolsep}{16pt}
\renewcommand{\arraystretch}{1.3}
\centering
\caption{Size of the SWB tests under correlated homoskedastic errors}
\label{tab:correrr_BALURT}
\vspace{.3cm}
\resizebox{\textwidth}{!}{
	
\begin{tabular}{l *{8}{S} }
\toprule
\toprule
\multicolumn{1}{c}{} & \multicolumn{4}{c}{$\vz_t=1$} & \multicolumn{4}{c}{$\vz_t=(1,t)'$} \\
\cmidrule(l{3pt}r{3pt}){2-5} \cmidrule(l{3pt}r{3pt}){6-9}
$T$ & $\tau^*$ & $\Breve{\tau}^*$ & $\tau$ & $\Breve{\tau}$ & $\tau^*$ & $\Breve{\tau}^*$ & $\tau$ & $\Breve{\tau}$\\
\midrule
\addlinespace[0.3em]
\multicolumn{9}{c}{\textbf{$\varphi = -.8,\, \vartheta = 0$}}\\
75 & 0.0466000 & 0.0422000 & 0.053600 & 0.0492000 & 0.0260 & 0.0328 & 0.0588 & 0.0678\\
100 & 0.0438000 & 0.0436000 & 0.050800 & 0.0468000 & 0.0300 & 0.0348 & 0.0512 & 0.0544\\
150 & 0.0414000 & 0.0400000 & 0.044600 & 0.0424000 & 0.0442 & 0.0468 & 0.0596 & 0.0596\\
250 & 0.0500000 & 0.0500000 & 0.050600 & 0.0488000 & 0.0436 & 0.0462 & 0.0502 & 0.0508\\
500 & 0.0520000 & 0.0488000 & 0.051000 & 0.0466000 & 0.0476 & 0.0520 & 0.0498 & 0.0514\\
1000 & 0.0498000 & 0.0524000 & 0.051400 & 0.0500000 & 0.0472 & 0.0472 & 0.0476 & 0.0448\\
\addlinespace[0.3em]
\multicolumn{9}{c}{\textbf{$\varphi = .8,\, \vartheta = 0$}}\\
75 & 0.0514000 & 0.0506000 & 0.056000 & 0.0396000 & 0.0542 & 0.0518 & 0.0454 & 0.0356\\
100 & 0.0466000 & 0.0494000 & 0.047200 & 0.0380000 & 0.0460 & 0.0460 & 0.0382 & 0.0320\\
150 & 0.0538000 & 0.0504000 & 0.051800 & 0.0396000 & 0.0462 & 0.0482 & 0.0412 & 0.0330\\
250 & 0.0490000 & 0.0470000 & 0.046000 & 0.0390000 & 0.0494 & 0.0468 & 0.0460 & 0.0356\\
500 & 0.0522000 & 0.0516000 & 0.050200 & 0.0446000 & 0.0490 & 0.0476 & 0.0434 & 0.0420\\
1000 & 0.0532000 & 0.0536000 & 0.052800 & 0.0476000 & 0.0532 & 0.0546 & 0.0492 & 0.0462\\
\addlinespace[0.3em]
\multicolumn{9}{c}{\textbf{$\varphi = 0,\, \vartheta = -.8$}}\\
75 & 0.1688000 & 0.1726000 & 0.316400 & 0.3250000 & 0.2770 & 0.3068 & 0.5002 & 0.5304\\
100 & 0.1280000 & 0.1364000 & 0.286600 & 0.2878000 & 0.1904 & 0.2246 & 0.4464 & 0.4746\\
150 & 0.0986000 & 0.0992000 & 0.253200 & 0.2510000 & 0.1242 & 0.1512 & 0.3996 & 0.4022\\
250 & 0.0874000 & 0.0848000 & 0.216000 & 0.2058000 & 0.0816 & 0.0910 & 0.3422 & 0.3274\\
500 & 0.0788000 & 0.0784000 & 0.170800 & 0.1688000 & 0.0690 & 0.0740 & 0.2870 & 0.2586\\
1000 & 0.0786000 & 0.0752000 & 0.135000 & 0.1362000 & 0.0800 & 0.0798 & 0.2384 & 0.2108\\
\addlinespace[0.3em]
\multicolumn{9}{c}{\textbf{$\varphi = 0,\, \vartheta = .8$}}\\
75 & 0.0420000 & 0.0404000 & 0.047600 & 0.0430000 & 0.0334 & 0.0368 & 0.0430 & 0.0424\\
100 & 0.0426000 & 0.0380000 & 0.051000 & 0.0392000 & 0.0330 & 0.0374 & 0.0448 & 0.0462\\
150 & 0.0434087 & 0.0430086 & 0.045209 & 0.0446089 & 0.0360 & 0.0340 & 0.0420 & 0.0456\\
250 & 0.0452000 & 0.0440000 & 0.049000 & 0.0492000 & 0.0432 & 0.0436 & 0.0508 & 0.0542\\
500 & 0.0548000 & 0.0548000 & 0.057600 & 0.0564000 & 0.0434 & 0.0436 & 0.0472 & 0.0502\\
1000 & 0.0486000 & 0.0506000 & 0.049800 & 0.0480000 & 0.0496 & 0.0460 & 0.0520 & 0.0494\\
\bottomrule
\end{tabular}
}
\begin{minipage}{\textwidth}
    \vspace{.25cm}
    \scriptsize\textit{Notes:} DGP \eqref{eq:ardgp_BALURT} with $c=0$ and $\sigma_t=1\,\forall$. The data are adjusted for a constant ($\vz_t = 1$) or a linear time trend ($\vz_t = (1,t)'$) using the FD method of \textcite{SchmidtPhillips1992}. All lag orders are computed using the RSMAIC. $B=499$ sieve wild bootstrap replications and 5000 Monte Carlo replications.
    \end{minipage}
\end{table}

\begin{table}[htbp]
\setlength{\tabcolsep}{16pt}
\renewcommand{\arraystretch}{1}
\centering
\caption{Size of the SWB activation knot tests under heteroskedastic MA errors}
\label{tab:hetero_correrr_BALURT}
\vspace{.3cm}
\resizebox{\textwidth}{!}{
	
\begin{tabular}{l *{8}{S} }
\toprule
\toprule
\multicolumn{1}{c}{} & \multicolumn{4}{c}{$\vz_t=1$} & \multicolumn{4}{c}{$\vz_t=(1,t)'$} \\
\cmidrule(l{3pt}r{3pt}){2-5} \cmidrule(l{3pt}r{3pt}){6-9}
$T$ & $\tau^*$ & $\Breve{\tau}^*$ & $\tau$ & $\Breve{\tau}$ & $\tau^*$ & $\Breve{\tau}^*$ & $\tau$ & $\Breve{\tau}$\\
\midrule
\addlinespace[.3em]
\multicolumn{9}{c}{\textit{Early smooth variance reduction, $\varphi = 0,\, \vartheta = 0$}}\\
75 & 0.0598 & 0.061000 & 0.0668 & 0.0786 & 0.0558 & 0.0576000 & 0.0808 & 0.0786\\
100 & 0.0576 & 0.061800 & 0.0628 & 0.0728 & 0.0520 & 0.0540000 & 0.0784 & 0.0760\\
150 & 0.0528 & 0.055600 & 0.0638 & 0.0750 & 0.0526 & 0.0554000 & 0.0812 & 0.0820\\
250 & 0.0496 & 0.052000 & 0.0628 & 0.0754 & 0.0558 & 0.0542000 & 0.0866 & 0.0860\\
500 & 0.0486 & 0.046000 & 0.0644 & 0.0732 & 0.0528 & 0.0496000 & 0.0868 & 0.0856\\
1000 & 0.0482 & 0.047600 & 0.0628 & 0.0754 & 0.0534 & 0.0564000 & 0.0832 & 0.0922\\
\addlinespace[.3em]
\multicolumn{9}{c}{\textit{Early smooth variance reduction, $\varphi = 0,\, \vartheta = -.8$}}\\
75 & 0.1504 & 0.184600 & 0.2694 & 0.3302 & 0.2730 & 0.3384677 & 0.5112 & 0.6050\\
100 & 0.1248 & 0.143600 & 0.2552 & 0.2982 & 0.2006 & 0.2610522 & 0.4734 & 0.5608\\
150 & 0.0946 & 0.109600 & 0.2192 & 0.2576 & 0.1340 & 0.1748000 & 0.4218 & 0.4834\\
250 & 0.0870 & 0.090400 & 0.2194 & 0.2334 & 0.0940 & 0.1174000 & 0.3702 & 0.3992\\
500 & 0.0744 & 0.071800 & 0.1874 & 0.2042 & 0.0838 & 0.0884000 & 0.3372 & 0.3336\\
1000 & 0.0826 & 0.076400 & 0.1642 & 0.1708 & 0.0756 & 0.0726000 & 0.2808 & 0.2702\\
\addlinespace[.3em]
\multicolumn{9}{c}{\textit{Early smooth variance reduction, $\varphi = 0,\, \vartheta = .8$}}\\
75 & 0.0588 & 0.064400 & 0.0818 & 0.0722 & 0.0564 & 0.0570000 & 0.1080 & 0.0892\\
100 & 0.0474 & 0.057000 & 0.0754 & 0.0720 & 0.0532 & 0.0604000 & 0.1098 & 0.0952\\
150 & 0.0508 & 0.058000 & 0.0724 & 0.0766 & 0.0482 & 0.0534000 & 0.0952 & 0.0918\\
250 & 0.0540 & 0.058400 & 0.0722 & 0.0820 & 0.0502 & 0.0510000 & 0.0936 & 0.0972\\
500 & 0.0512 & 0.055200 & 0.0724 & 0.0884 & 0.0518 & 0.0534000 & 0.0976 & 0.1066\\
1000 & 0.0556 & 0.056200 & 0.0684 & 0.0836 & 0.0470 & 0.0486000 & 0.0854 & 0.0976\\
\addlinespace[.3em]
\multicolumn{9}{c}{\textit{Late smooth variance increase, $\varphi = 0,\, \vartheta = 0$}}\\
75 & 0.0636 & 0.056200 & 0.1364 & 0.1082 & 0.0554 & 0.0582000 & 0.0730 & 0.0766\\
100 & 0.0624 & 0.061600 & 0.1538 & 0.1294 & 0.0484 & 0.0532000 & 0.0662 & 0.0694\\
150 & 0.0562 & 0.055200 & 0.1494 & 0.1186 & 0.0512 & 0.0514000 & 0.0774 & 0.0810\\
250 & 0.0548 & 0.051200 & 0.1522 & 0.1210 & 0.0500 & 0.0478000 & 0.0778 & 0.0802\\
500 & 0.0552 & 0.050600 & 0.1556 & 0.1226 & 0.0516 & 0.0540000 & 0.0860 & 0.0880\\
1000 & 0.0470 & 0.049000 & 0.1478 & 0.1254 & 0.0502 & 0.0572000 & 0.0896 & 0.0958\\
\addlinespace[.3em]
\multicolumn{9}{c}{\textit{Late smooth variance increase, $\varphi = 0,\, \vartheta = -.8$}}\\
75 & 0.2062 & 0.214200 & 0.4172 & 0.4240 & 0.1902 & 0.2264000 & 0.3114 & 0.3764\\
100 & 0.1622 & 0.175035 & 0.3848 & 0.3894 & 0.1496 & 0.1854000 & 0.2884 & 0.3516\\
150 & 0.1316 & 0.137000 & 0.3518 & 0.3656 & 0.0954 & 0.1300000 & 0.2632 & 0.3138\\
250 & 0.1082 & 0.109000 & 0.3194 & 0.3370 & 0.0694 & 0.0898000 & 0.2426 & 0.2878\\
500 & 0.0876 & 0.087800 & 0.2844 & 0.2886 & 0.0686 & 0.0714000 & 0.2410 & 0.2594\\
1000 & 0.0802 & 0.082600 & 0.2476 & 0.2584 & 0.0658 & 0.0660000 & 0.2266 & 0.2412\\
\addlinespace[.3em]
\multicolumn{9}{c}{\textit{Late smooth variance increase, $\varphi = 0,\, \vartheta = .8$}}\\
75 & 0.0636 & 0.065600 & 0.1572 & 0.1322 & 0.0470 & 0.0510000 & 0.0548 & 0.0870\\
100 & 0.0654 & 0.061600 & 0.1672 & 0.1376 & 0.0474 & 0.0492000 & 0.0590 & 0.0900\\
150 & 0.0604 & 0.056600 & 0.1544 & 0.1238 & 0.0512 & 0.0538000 & 0.0672 & 0.0978\\
250 & 0.0662 & 0.061600 & 0.1718 & 0.1358 & 0.0502 & 0.0492000 & 0.0694 & 0.0960\\
500 & 0.0588 & 0.051200 & 0.1556 & 0.1268 & 0.0584 & 0.0548000 & 0.0844 & 0.1060\\
1000 & 0.0540 & 0.058000 & 0.1660 & 0.1338 & 0.0552 & 0.0554000 & 0.0790 & 0.0950\\
\bottomrule
\end{tabular}
}
\begin{minipage}{\textwidth}
    \vspace{.25cm}
    \scriptsize\textit{Notes:} 5\% nominal level. DGP \eqref{eq:ardgp_BALURT} with $c=0$ and $\sigma_{T,\,t}$ as defined in \eqref{eq:svt_BALURT}, with $\kappa=.2$, $s_2^2 = .25$ for early negative shifts and $\kappa=.8$, $s_2^2=4$ for late positive shifts. The data are adjusted for a constant ($\vz_t = 1$) or a linear time trend ($\vz_t = (1,t)'$) using the FD method of \textcite{SchmidtPhillips1992}. All lag orders are selected using the RSMAIC. $B=499$ sieve wild bootstrap replications. $5000$ Monte Carlo replications.
    \end{minipage}
\end{table}

Next, we investigate the impact of unconditional heteroscedasticity and its interplay with the effects of autocorrelated errors on the precision of the tests. Our focus is MA processes for which $\tau$ and $\Breve{\tau}$ show the largest size distortions, and bootstrap inference with better precision is the most needed. \Cref{tab:hetero_correrr_BALURT} presents the size of the tests for MA errors with non-stationary volatility. Consistent with the theory, we see (upward) size distortions for $\tau$ and $\Breve{\tau}$ under smoothly trending variances across all scenarios. These distortions vary in magnitude with the correlation structure, the type of variance shift and the adjustment for $\vz_t$. The smallest extent occurs for an early variance reduction and MA coefficient $\vartheta = .8$, where both tests mostly remain below 10\%. There are differences between $\tau$ and $\Breve{\tau}$, e.g., for a late variance increase with $\vartheta = .8$, which is likely due to idiosyncracies of the $J_\alpha$ statistic under heteroskedasticity. As before, we observe the most pronounced distortions at $\vartheta=-.8$ for small $T$, which are worse than under homoscedasticity, cf. \Cref{tab:correrr_BALURT}. The bootstrap analogues consistently perform better, having size close to the 5\% level for positive MA coefficients for upward and downward variance shifts and independently of the adjustment for a deterministic component. For $\vartheta=-.8$, $\tau^*$ and $\Breve{\tau}^*$ tend to be somewhat more oversized than under homoscedasticity and also require large $T$ for rejection rates to ameliorate. As in the homoskedastic settings, comparing with additional simulation results for the WB tests with $q=0$ in \Cref{tab:tab_TA6_BALURT} indicates better precision of the SWB tests. \Cref{tab:tab_TA1_BALURT,tab:tab_TA7_BALURT} report qualitatively similar results for heteroskedastic AR errors.

Local power estimates for the SWB tests under heteroskedastic AR and MA processes are provided in \Cref{tab:tab_TA3_BALURT,tab:tab_TA4_BALURT}. With a few exceptions for error processes with negative coefficients and small $T$, we find that the bootstrap tests effectively approximate the local power functions of the infeasible size-adjusted tests $\tau$ and $\Breve{\tau}$. 
Furthermore, the simulation results show that information enrichment yields power gains even under non-stationary volatility and that the bootstrap retains these gains.

\section{Application to Real Residential Property Prices}
\label{sec:ea_BALURT}
\begin{figure}[H]
\centering
\caption{RRPPI rates and estimated variance profiles of the top six Euro area economies}
\label{fig:rrpinf_plot}
\vspace{.25cm}
\begin{minipage}{.49\textwidth}
	\includegraphics[width=\textwidth]{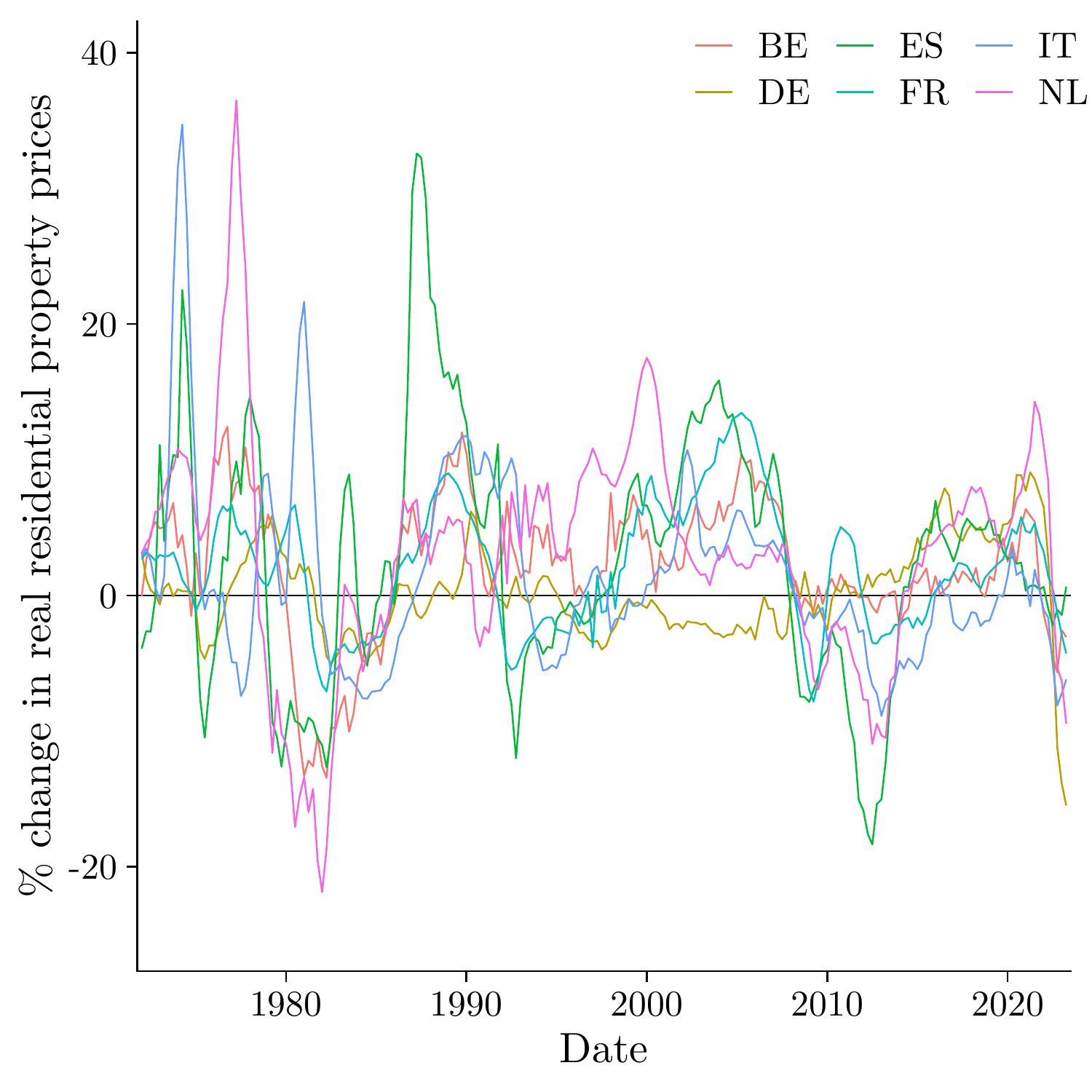}	
\end{minipage}
\begin{minipage}{.49\textwidth}
		\includegraphics[width=\textwidth]{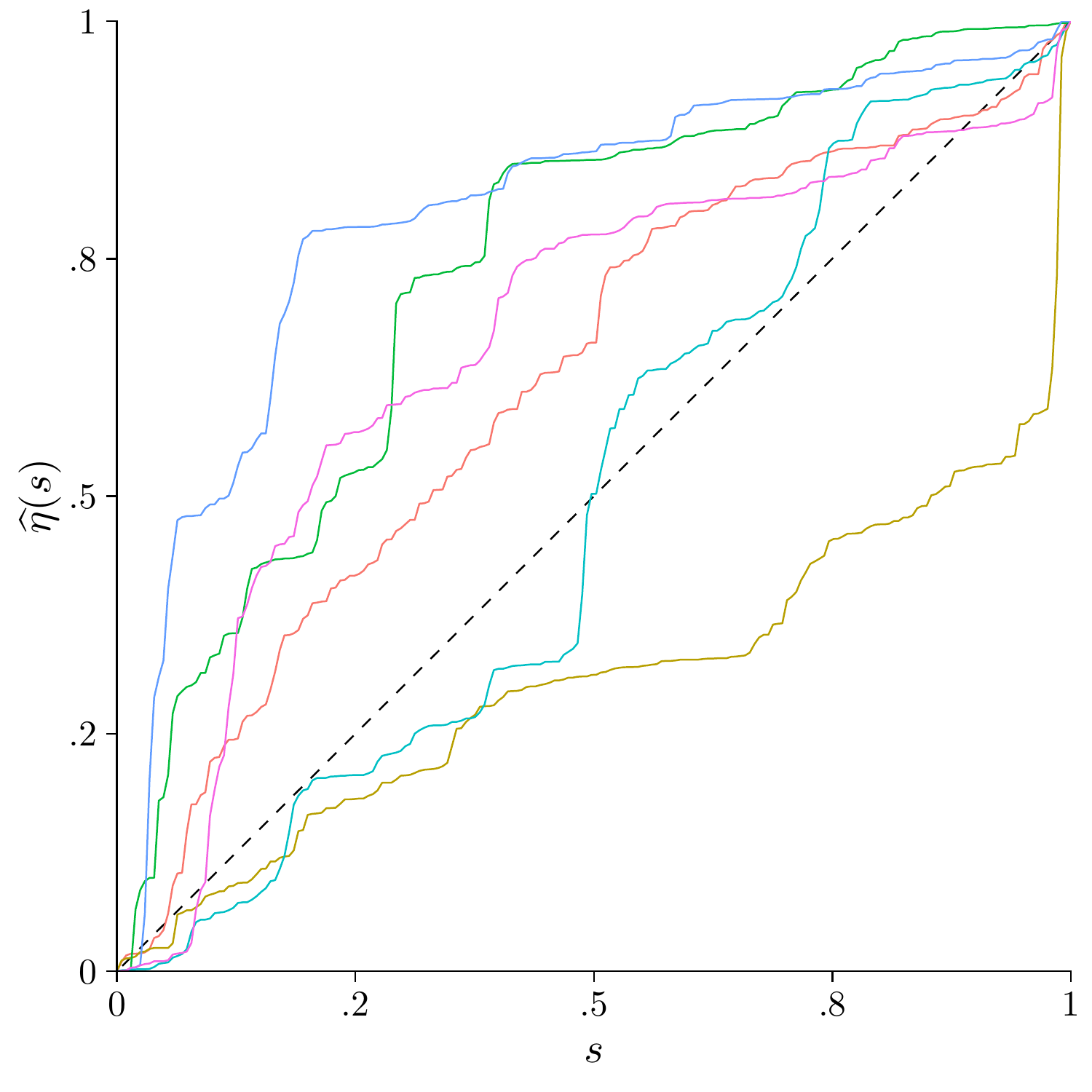}	
\end{minipage}
\vspace{.25cm}
\begin{minipage}{\textwidth}
	\scriptsize\textit{Notes:} Left: Year-on-year quarterly real residential property price inflation rates of selected OECD countries from Q1-1972 to Q2-2023 ($T=207$). Right: estimated variance profiles, computed as $\widehat{\eta}(s) := \left(\sum_{t=1}^{\lfloor sT\rfloor}\widehat{u}_t^2 + (sT - \lfloor sT \rfloor)\widehat{u}_{\lfloor sT\rfloor + 1}\right) \big/ \sum_{t=1}^T \widehat{u}_t^2$, with $\widehat{u}_t$ the residuals from an AR(1) OLS regression in levels, cf. \textcite{CavaliereTaylor2007}.
	\end{minipage}
\end{figure}

The housing price rally sparked by a sustained low-interest environment after the 2008 financial crisis has received much public and scientific attention \parencite[see, e.g.,][]{Mianetal2015,Jordaetal2016}. While monetary policies fueled price dynamics during the COVID-19 pandemic \parencite[cf.][]{FranckeKorevaar2021}, the recent trend reversal due to surges in mortgage and construction interest rates highlights the importance of understanding the dynamics in the market relevant to macroeconomic policy.

We use the tests to assess real residential property price inflation (RRPPI) rates of the six largest Eurozone economies for stochastic trends, covering 207 quarterly observations from Q1-1972 to Q2-2023 for Belgium, France, Germany, Italy, Spain and the Netherlands. The data \parencite{RRPPdata} obtained from CPI deflated price indices summarise all types of new and existing dwellings, except for Germany, where the underlying index considers owner-occupied houses only.

Besides the abovementioned events, the sample period covers further economic turmoil, such as OPEC oil shortages (1973, 1979) and recessions in the 1980s and 1990s, reflecting different variance regimes. We find evidence of such regimes in data displayed in the left panel of \Cref{fig:rrpinf_plot}, showcasing a high degree of co-movement. The estimated volatility profiles in the right panel further indicate the instability of the unconditional variances, showing considerable downward shifts for Belgium (BE), Italy (IT), Spain (ES), and the Netherlands (NL).

\begin{table}[t]
\centering
\caption{Test outcomes for European residential property price inflation rates}
\label{tab:rrppinfoutcomes}
\vspace{.25cm}
\setlength{\tabcolsep}{20pt}
\renewcommand{\arraystretch}{1}
\resizebox{\textwidth}{!}{
\begin{tabular}{lp{25pt} *{6}{S}}
\toprule
\toprule
&& {BE} & {DE} & {ES} & {FR} & {IT} & {NL} \\
\midrule
\addlinespace[0.5em]
\multicolumn{8}{l}{\textit{Entire sample: Q1-1972 -- Q2-2023} ($T = 206$)}\\[.5em]
&$\widehat{k}_i$ & {12} & {12} & {5} & {8} & {5} & {8}\\[.3em]
&$\tau$ & 0.00681399999999999$^{***}$ & 0.230492$^{}$ & 0.016222$^{**}$ & 0.011808$^{**}$ & {<.001}$^{***}$ & 0.00243199999999999$^{***}$\\
&$\Breve{\tau}$ & {<} .001$^{***}$ & 0.056754$^{*}$ & 0.00189600000000001$^{***}$ & 0.00796600000000003$^{***}$ & {<\,.001}$^{***}$ & 0.00318399999999996$^{***}$\\[.3em]
&$\tau^*$ & 0.00980196039207842$^{***}$ & 0.298059611922385$^{}$ & 0.02000400080016$^{**}$ & 0.0162032406481296$^{**}$ & {<.001$^{***}$} & 0.00480096019203841$^{***}$\\ 
&$\Breve{\tau}^*$ & {<}.001$^{***}$ & 0.107221444288858$^{}$ & 0.0016003200640128$^{***}$ & 0.011002200440088$^{**}$ & {<\,.001}$^{***}$ & 0.00620124024804961$^{***}$\\
\addlinespace[1em]

\multicolumn{8}{l}{\textit{Pre-Euro: Q1-1972 -- Q4-1998} ($T = 108$)}\\[.5em]
&$\widehat{k}_i$ & {8} & {4} & {4} & {4} & {2} & {4}\\[.3em]
&$\tau$ & 0.027678$^{**}$ & 0.378466$^{}$ & 0.109232$^{}$ & 0.066306$^{*}$ & {<\,.001}$^{***}$ & 0.032158$^{**}$\\
&$\Breve{\tau}$ & 0.0201480000000001$^{**}$ & 0.16953$^{}$ & 0.056536$^{*}$ & 0.048574$^{**}$ & {<\,.001}$^{***}$ & 0.036956$^{**}$\\[.3em]
&$\tau^*$ & 0.0442088417683537$^{**}$ & 0.359071814362873$^{}$ & 0.0856171234246849$^{*}$ & 0.100220044008802$^{}$ & {<.001}$^{***}$ & 0.0268053610722144$^{**}$\\
&$\Breve{\tau}^*$ & 0.0146029205841168$^{**}$ & 0.175635127025405$^{}$ & 0.0524104820964193$^{*}$ & 0.0514102820564113$^{*}$ & {<.001}$^{***}$ & 0.0308061612322464$^{**}$\\
\addlinespace[1em]

\multicolumn{8}{l}{\textit{Euro-era: Q1-1999 -- Q2-2023} ($T = 98$)}\\[.5em]
&$\widehat{k}_i$ & {4} & {8} & {4} & {8} & {4} & {4}\\[.3em]
&$\tau$ & 0.569222$^{}$ & 0.107582$^{}$ & 0.242626$^{}$ & 0.38567$^{}$ & 0.269016$^{}$ & 0.61541$^{}$\\
&$\Breve{\tau}$ & 0.356342$^{}$ & 0.051964$^{*}$ & 0.193808$^{}$ & 0.289602$^{}$ & 0.212922$^{}$ & 0.476674$^{}$\\[.3em]
&$\tau^*$ & 0.588917783556711 & 0.197039407881576 & 0.256651330266053 & 0.43868773754751 & 0.274654930986197 & 0.673134626925385\\
&$\Breve{\tau}^*$ & 0.470694138827766 & 0.12122424484897 & 0.213842768553711 & 0.406081216243249 & 0.2750550110022 & 0.609721944388878\\
\addlinespace[0.5em]

\bottomrule
\addlinespace[.3em]
\end{tabular}
}
\begin{minipage}{\textwidth}
        \vspace{.1em}
        \scriptsize\textit{Notes:} quarterly year-on-year RRPPI rates from \textcite{RRPPdata}. The tests are computed on FD-demeaned data, and the RSMAIC selected all lag orders with maximum lag order $k_{\max}=\lfloor12\cdot(100/T)^{.25}\rfloor$. $\widehat{k}_i$ is the estimated truncation lag in model \eqref{eq:eareg_BALURT}. The test outcomes are p-values. The supercripts $^{*}$, $^{**}$ and $^{***}$ indicate a rejection of the unit root null at 10\%, 5\% and 1\%, respectively. The bootstrap tests $\tau$ and $\Breve{\tau}$ are computed based on $B=4999$ bootstrap replications.
        \end{minipage}
\end{table}

We consider ADF regressions with a non-zero mean,
\begin{align}
	\Delta \text{RRPPI}_{i,\,t} = \mu_i + \beta_0\, \text{RRPPI}_{i,\,t-1} + \sum_{j=1}^{k_i} \beta_j\, \Delta\text{RRPPI}_{i,\,t-j} + e_{i,\,t},\label{eq:eareg_BALURT}
\end{align}
where the $k_i$ in the baseline regression, as well as lag orders in \Cref{algo:wbftau}, are selected by RSMAIC. We compute the bootstrap tests using $B=4999$ bootstrap iterations.

\Cref{tab:rrppinfoutcomes} presents the outcomes of the tests for three different periods: the entire data range (Q1-1972--Q2-2023) and subsamples before (Q1-1972--Q4-1998) and for the Euro-era (Q1-1999--Q2-2023) which is characterised by macroeconomic convergence, e.g., due to a mutual primary refinancing interest rate applying to the member economies. 

Outcomes for the bootstrap tests mostly agree with the standard inference, giving mixed conclusions. For the entire period, the tests indicate mean reversion behaviour for all economies except Germany, where only $\breve{\tau}$ rejects at 5\%. The pre-Euro subsample shows similar outcomes. Notably, $\breve{\tau}^*$ tends to have smaller p-values than $\tau^*$, likely due to power gains from information enrichment—a feature of $\breve{\tau}$ that the simulation outcomes indicate is preserved by the bootstrap. None of the tests rejects the null hypothesis of a stochastic trend for the Euro-era at $5\%$. The reduced sample size or possible trend changes in the generating process could contribute to the mixed evidence, especially since periods of exuberance in the housing market characterise the Euro-era subsample.

\section{Conclusion}
\label{sec:conclusion_BALURT}
In this paper, we assessed the reliability of the adaptive Lasso solution path-based unit root tests $\tau$ and $\Breve{\tau}$ proposed in \textcite{Arnold2024} under weaker assumptions and considered whether resampling offers robust alternatives. Drawing on the theoretical results in \textcite{CavaliereTaylor2008,CavaliereTaylor2009}, we propose the wild bootstrap analogues $\tau^*$ and $\Breve{\tau}^*$, implementing the computation of resampled Lasso solution paths efficiently using the LARS algorithm \parencite{Efronetal2004, pkg-lars}. Numerical evidence shows that the bootstrap yields tests with higher precision, allowing more robust inference under correlated error processes of a general form. 

Consistent with the theory on heteroskedastic autoregressions in \textcite{CavaliereTaylor2007}, $\tau$ and $\Breve{\tau}$ do not attain their homoscedastic limits derived in \textcite{Arnold2024} when the errors are unconditionally heteroscedastic, so that valid inference is not guaranteed, even asymptotically. Our simulations confirm that $\tau$ and $\Breve{\tau}$---like conventional unit root tests---have null distributions and local power functions affected by nuisance parameters for unconditionally heteroskedastic innovations. A consequence is size distortions, with the strength of the effect depending on the adjustment for deterministic components. Correlated errors, such as MA(1) processes with negative coefficients, exacerbate this effect. Our sieve wild bootstrap tests display higher precision than $\tau$ and $\Breve{\tau}$ in these scenarios, indicating their ability to accurately recover the first-order null distributions. In addition, our simulation results show the wild bootstrap variants to approximate the finite-sample (local) power functions of the infeasible size-adjusted implementations of the standard tests under variance shifts.

To illustrate the bootstrap tests, we consider real residential property price inflation rates for selected Eurozone economies. This data set seems representative of our setup, given signs of persistence and non-stationary volatility. Both bootstrap tests yield the same conclusions, pointing to stationarity for the entire period from 1972 to 2023 and the period before the introduction of the Euro. We find no evidence for mean-reversion for the Euro-era subsample.

There are various avenues for further research. Our simulation evidence suggests that the wild bootstrap preserves the benefits of information enrichment, motivating its application to inference for penalised regression when the (asymptotic) distribution is unknown.
Given the positive results for $\tau$ and $\Breve{\tau}$, expanding the underlying testing principle to other (adaptively) penalised regression estimators, for example, the fused Lasso or the group Lasso seems worthwhile. Good starting points are \textcite{Qian2016} and \textcite{Schweikert2021}, which employ adaptive variants of the group Lasso and fused Lasso to detect structural breaks in panel and cointegrating regressions. 

Our empirical application raises the question of whether the advent of the Euro affected the heterogeneity in the housing price dynamics across the Eurozone countries. To address this question, contemplating additional measures of heterogeneity for information enrichment can improve the discriminatory power of the aforementioned penalised estimators. This topic is closely related to the heterogeneous treatment effect inference literature, which could provide further inspiration for extensions.

\enlargethispage{\baselineskip} 
Another promising direction is inference in high-dimensional regressions for which the double Lasso of \textcite{Bellonietal2014} has become a standard method. It would be appealing to investigate whether information enrichment is still adequate in high-dimensional regression inference for which the wild bootstrap has become a cornerstone \parencite[cf.][]{Chernozhukov2023}. To this end, one could apply our testing principle to causal inference problems or multivariate time series regressions, potentially using adaptive penalty weights derived using the de-sparsified Lasso. We are currently investigating this approach.
\clearpage
\appendix
\addcontentsline{toc}{section}{Appendix}
\renewcommand{\thesubsection}{\Alph{subsection}}
\renewcommand{\theequation}{\thesubsection\arabic{equation}}
\setcounter{table}{0}
\setcounter{figure}{0}
\renewcommand{\thetable}{\thesubsection\arabic{table}}
\renewcommand{\thefigure}{\thesubsection\arabic{figure}}
%

\subsection{Additional Simulation Results}
\label[appendix]{sec:asr_BALURT}

\begin{table}[ht]
\setlength{\tabcolsep}{16pt}
\renewcommand{\arraystretch}{1.1}
\centering
\caption{Size of the WB activation knot tests under correlated homoskedastic errors}
\label{tab:tab_TA5_BALURT}
\vspace{.3cm}
\resizebox{\textwidth}{!}{
	
\begin{tabular}{l *{8}{S} }
\toprule
\toprule
\multicolumn{1}{c}{} & \multicolumn{4}{c}{$\vz_t=1$} & \multicolumn{4}{c}{$\vz_t=(1,t)'$} \\
\cmidrule(l{3pt}r{3pt}){2-5} \cmidrule(l{3pt}r{3pt}){6-9}
$T$ & $\tau^*$ & $\Breve{\tau}^*$ & $\tau$ & $\Breve{\tau}$ & $\tau^*$ & $\Breve{\tau}^*$ & $\tau$ & $\Breve{\tau}$\\
\midrule
\addlinespace[0.3em]
\multicolumn{9}{c}{\textbf{$\varphi = -.8,\, \vartheta = 0$}}\\
75 & 0.0566000 & 0.0552000 & 0.053600 & 0.0492000 & 0.0634 & 0.0726 & 0.0588 & 0.0678\\
100 & 0.0532000 & 0.0552000 & 0.050800 & 0.0468000 & 0.0544 & 0.0600 & 0.0512 & 0.0544\\
150 & 0.0492000 & 0.0484000 & 0.044600 & 0.0424000 & 0.0606 & 0.0666 & 0.0596 & 0.0596\\
250 & 0.0560000 & 0.0562000 & 0.050600 & 0.0488000 & 0.0526 & 0.0566 & 0.0502 & 0.0508\\
500 & 0.0534000 & 0.0534000 & 0.051000 & 0.0466000 & 0.0518 & 0.0564 & 0.0498 & 0.0514\\
1000 & 0.0548000 & 0.0554000 & 0.051400 & 0.0500000 & 0.0510 & 0.0528 & 0.0476 & 0.0448\\
\addlinespace[0.3em]
\multicolumn{9}{c}{\textbf{$\varphi = .8,\, \vartheta = 0$}}\\
75 & 0.0596000 & 0.0422000 & 0.056000 & 0.0396000 & 0.0472 & 0.0392 & 0.0454 & 0.0356\\
100 & 0.0522000 & 0.0430000 & 0.047200 & 0.0380000 & 0.0396 & 0.0380 & 0.0382 & 0.0320\\
150 & 0.0588000 & 0.0446000 & 0.051800 & 0.0396000 & 0.0450 & 0.0388 & 0.0412 & 0.0330\\
250 & 0.0500000 & 0.0440000 & 0.046000 & 0.0390000 & 0.0472 & 0.0434 & 0.0460 & 0.0356\\
500 & 0.0544000 & 0.0508000 & 0.050200 & 0.0446000 & 0.0454 & 0.0466 & 0.0434 & 0.0420\\
1000 & 0.0570000 & 0.0552000 & 0.052800 & 0.0476000 & 0.0512 & 0.0522 & 0.0492 & 0.0462\\
\addlinespace[0.3em]
\multicolumn{9}{c}{\textbf{$\varphi = 0,\, \vartheta = -.8$}}\\
75 & 0.3316000 & 0.3416000 & 0.316400 & 0.3250000 & 0.5140 & 0.5494 & 0.5002 & 0.5304\\
100 & 0.2964000 & 0.3022000 & 0.286600 & 0.2878000 & 0.4574 & 0.4932 & 0.4464 & 0.4746\\
150 & 0.2612000 & 0.2638000 & 0.253200 & 0.2510000 & 0.4104 & 0.4202 & 0.3996 & 0.4022\\
250 & 0.2210000 & 0.2168000 & 0.216000 & 0.2058000 & 0.3530 & 0.3462 & 0.3422 & 0.3274\\
500 & 0.1774000 & 0.1782000 & 0.170800 & 0.1688000 & 0.2918 & 0.2694 & 0.2870 & 0.2586\\
1000 & 0.1374000 & 0.1402000 & 0.135000 & 0.1362000 & 0.2448 & 0.2214 & 0.2384 & 0.2108\\
\addlinespace[0.3em]
\multicolumn{9}{c}{\textbf{$\varphi = 0,\, \vartheta = .8$}}\\
75 & 0.0590000 & 0.0516000 & 0.047600 & 0.0430000 & 0.0478 & 0.0514 & 0.0430 & 0.0424\\
100 & 0.0616000 & 0.0488000 & 0.051000 & 0.0392000 & 0.0496 & 0.0586 & 0.0448 & 0.0462\\
150 & 0.0522104 & 0.0506101 & 0.045209 & 0.0446089 & 0.0468 & 0.0546 & 0.0420 & 0.0456\\
250 & 0.0564000 & 0.0574000 & 0.049000 & 0.0492000 & 0.0552 & 0.0626 & 0.0508 & 0.0542\\
500 & 0.0630000 & 0.0618000 & 0.057600 & 0.0564000 & 0.0522 & 0.0588 & 0.0472 & 0.0502\\
1000 & 0.0528000 & 0.0534000 & 0.049800 & 0.0480000 & 0.0552 & 0.0554 & 0.0520 & 0.0494\\
\bottomrule
\end{tabular}
}
\begin{minipage}{\textwidth}
    \vspace{.25cm}
    \scriptsize\textit{Notes:} DGP \eqref{eq:ardgp_BALURT} with $c=0$ and $\sigma_{T,\,t}$ as defined in \eqref{eq:svt_BALURT}, with $\kappa=.2$, $s_2^2 = .25$ for early negative shifts and $\kappa=.8$, $s_2^2=4$ for late positive shifts. The data are adjusted for a constant ($\vz_t = 1$) or a linear time trend ($\vz_t = (1,t)'$) using the FD method of \textcite{SchmidtPhillips1992}. All lag orders are computed using the RSMAIC. $B=499$ wild bootstrap replications with $q=0$. $5000$ Monte Carlo replications.
    \end{minipage}
\end{table}

\begin{table}[htbp]
\setlength{\tabcolsep}{16pt}
\renewcommand{\arraystretch}{1}
\centering
\caption{Size of the WB activation knot tests under heteroskedastic MA errors}
\label{tab:tab_TA6_BALURT}
\vspace{.3cm}
\resizebox{\textwidth}{!}{
	
\begin{tabular}{l *{8}{S} }
\toprule
\toprule
\multicolumn{1}{c}{} & \multicolumn{4}{c}{$\vz_t=1$} & \multicolumn{4}{c}{$\vz_t=(1,t)'$} \\
\cmidrule(l{3pt}r{3pt}){2-5} \cmidrule(l{3pt}r{3pt}){6-9}
$T$ & $\tau^*$ & $\Breve{\tau}^*$ & $\tau$ & $\Breve{\tau}$ & $\tau^*$ & $\Breve{\tau}^*$ & $\tau$ & $\Breve{\tau}$\\
\midrule
\addlinespace[.3em]
\multicolumn{9}{c}{\textit{Early smooth variance reduction, $\varphi = 0,\, \vartheta = 0$}}\\
75 & 0.0650 & 0.0636 & 0.0668 & 0.0786 & 0.0592 & 0.0588 & 0.0808 & 0.0786\\
100 & 0.0600 & 0.0614 & 0.0628 & 0.0728 & 0.0544 & 0.0552 & 0.0784 & 0.0760\\
150 & 0.0558 & 0.0566 & 0.0638 & 0.0750 & 0.0578 & 0.0584 & 0.0812 & 0.0820\\
250 & 0.0524 & 0.0540 & 0.0628 & 0.0754 & 0.0584 & 0.0560 & 0.0866 & 0.0860\\
500 & 0.0490 & 0.0470 & 0.0644 & 0.0732 & 0.0544 & 0.0518 & 0.0868 & 0.0856\\
1000 & 0.0484 & 0.0476 & 0.0628 & 0.0754 & 0.0546 & 0.0586 & 0.0832 & 0.0922\\
\addlinespace[.3em]
\multicolumn{9}{c}{\textit{Early smooth variance reduction, $\varphi = 0,\, \vartheta = -.8$}}\\
75 & 0.2708 & 0.3222 & 0.2694 & 0.3302 & 0.4864 & 0.5814 & 0.5112 & 0.6050\\
100 & 0.2526 & 0.2844 & 0.2552 & 0.2982 & 0.4456 & 0.5304 & 0.4734 & 0.5608\\
150 & 0.2168 & 0.2412 & 0.2192 & 0.2576 & 0.3878 & 0.4482 & 0.4218 & 0.4834\\
250 & 0.2134 & 0.2110 & 0.2194 & 0.2334 & 0.3402 & 0.3598 & 0.3702 & 0.3992\\
500 & 0.1760 & 0.1714 & 0.1874 & 0.2042 & 0.2966 & 0.2866 & 0.3372 & 0.3336\\
1000 & 0.1502 & 0.1406 & 0.1642 & 0.1708 & 0.2372 & 0.2214 & 0.2808 & 0.2702\\
\addlinespace[.3em]
\multicolumn{9}{c}{\textit{Early smooth variance reduction, $\varphi = 0,\, \vartheta = .8$}}\\
75 & 0.0748 & 0.0574 & 0.0818 & 0.0722 & 0.0824 & 0.0632 & 0.1080 & 0.0892\\
100 & 0.0704 & 0.0554 & 0.0754 & 0.0720 & 0.0848 & 0.0670 & 0.1098 & 0.0952\\
150 & 0.0646 & 0.0590 & 0.0724 & 0.0766 & 0.0722 & 0.0670 & 0.0952 & 0.0918\\
250 & 0.0622 & 0.0610 & 0.0722 & 0.0820 & 0.0658 & 0.0656 & 0.0936 & 0.0972\\
500 & 0.0586 & 0.0610 & 0.0724 & 0.0884 & 0.0634 & 0.0672 & 0.0976 & 0.1066\\
1000 & 0.0556 & 0.0588 & 0.0684 & 0.0836 & 0.0532 & 0.0582 & 0.0854 & 0.0976\\
\addlinespace[.3em]
\multicolumn{9}{c}{\textit{Late smooth variance increase, $\varphi = 0,\, \vartheta = 0$}}\\
75 & 0.0656 & 0.0576 & 0.1364 & 0.1082 & 0.0616 & 0.0642 & 0.0730 & 0.0766\\
100 & 0.0688 & 0.0664 & 0.1538 & 0.1294 & 0.0540 & 0.0564 & 0.0662 & 0.0694\\
150 & 0.0586 & 0.0552 & 0.1494 & 0.1186 & 0.0538 & 0.0552 & 0.0774 & 0.0810\\
250 & 0.0568 & 0.0532 & 0.1522 & 0.1210 & 0.0530 & 0.0518 & 0.0778 & 0.0802\\
500 & 0.0564 & 0.0514 & 0.1556 & 0.1226 & 0.0534 & 0.0562 & 0.0860 & 0.0880\\
1000 & 0.0472 & 0.0490 & 0.1478 & 0.1254 & 0.0512 & 0.0570 & 0.0896 & 0.0958\\
\addlinespace[.3em]
\multicolumn{9}{c}{\textit{Late smooth variance increase, $\varphi = 0,\, \vartheta = -.8$}}\\
75 & 0.3512 & 0.3864 & 0.4172 & 0.4240 & 0.3032 & 0.3680 & 0.3114 & 0.3764\\
100 & 0.3110 & 0.3454 & 0.3848 & 0.3894 & 0.2758 & 0.3400 & 0.2884 & 0.3516\\
150 & 0.2782 & 0.3192 & 0.3518 & 0.3656 & 0.2482 & 0.2990 & 0.2632 & 0.3138\\
250 & 0.2310 & 0.2720 & 0.3194 & 0.3370 & 0.2240 & 0.2634 & 0.2426 & 0.2878\\
500 & 0.1786 & 0.2202 & 0.2844 & 0.2886 & 0.2084 & 0.2240 & 0.2410 & 0.2594\\
1000 & 0.1412 & 0.1824 & 0.2476 & 0.2584 & 0.1896 & 0.2002 & 0.2266 & 0.2412\\
\addlinespace[.3em]
\multicolumn{9}{c}{\textit{Late smooth variance increase, $\varphi = 0,\, \vartheta = .8$}}\\
75 & 0.0828 & 0.0810 & 0.1572 & 0.1322 & 0.0432 & 0.0754 & 0.0548 & 0.0870\\
100 & 0.0812 & 0.0776 & 0.1672 & 0.1376 & 0.0508 & 0.0756 & 0.0590 & 0.0900\\
150 & 0.0698 & 0.0730 & 0.1544 & 0.1238 & 0.0504 & 0.0790 & 0.0672 & 0.0978\\
250 & 0.0716 & 0.0734 & 0.1718 & 0.1358 & 0.0498 & 0.0706 & 0.0694 & 0.0960\\
500 & 0.0632 & 0.0620 & 0.1556 & 0.1268 & 0.0560 & 0.0674 & 0.0844 & 0.1060\\
1000 & 0.0570 & 0.0638 & 0.1660 & 0.1338 & 0.0542 & 0.0622 & 0.0790 & 0.0950\\
\bottomrule
\end{tabular}
}
\begin{minipage}{\textwidth}
    \vspace{.25cm}
    \scriptsize\textit{Notes:} 5\% nominal level. DGP \eqref{eq:ardgp_BALURT} with $c=0$ and $\sigma_{T,\,t}$ as defined in \eqref{eq:svt_BALURT}, with $\kappa=.2$, $s_2^2 = .25$ for early negative shifts and $\kappa=.8$, $s_2^2=4$ for late positive shifts. The data are adjusted for a constant ($\vz_t = 1$) or a linear time trend ($\vz_t = (1,t)'$) using the FD method of \textcite{SchmidtPhillips1992}. All lag orders are selected using the RSMAIC. $B=499$ wild bootstrap replications with $q=0$. $5000$ Monte Carlo replications.
    \end{minipage}
\end{table}

\begin{table}[ht]
\setlength{\tabcolsep}{16pt}
\renewcommand{\arraystretch}{1.1}
\centering
\caption{Size of the WB activation knot tests under heteroskedastic AR errors}
\label{tab:tab_TA7_BALURT}
\vspace{.3cm}
\resizebox{\textwidth}{!}{
	
\begin{tabular}{l *{8}{S} }
\toprule
\toprule
\multicolumn{1}{c}{} & \multicolumn{4}{c}{$\vz_t=1$} & \multicolumn{4}{c}{$\vz_t=(1,t)'$} \\
\cmidrule(l{3pt}r{3pt}){2-5} \cmidrule(l{3pt}r{3pt}){6-9}
$T$ & $\tau^*$ & $\Breve{\tau}^*$ & $\tau$ & $\Breve{\tau}$ & $\tau^*$ & $\Breve{\tau}^*$ & $\tau$ & $\Breve{\tau}$\\
\midrule
\addlinespace[.3em]
\multicolumn{9}{c}{\textit{Early smooth variance reduction, AR errors: $\varphi = -.8,\, \vartheta = 0$}}\\
75 & 0.0882000 & 0.0838000 & 0.0900000 & 0.0892000 & 0.1320 & 0.1196 & 0.1548 & 0.1346\\
100 & 0.0870000 & 0.0816000 & 0.0914000 & 0.0880000 & 0.1044 & 0.1064 & 0.1264 & 0.1232\\
150 & 0.0762000 & 0.0734000 & 0.0814000 & 0.0834000 & 0.0838 & 0.0878 & 0.1096 & 0.1090\\
250 & 0.0710000 & 0.0712000 & 0.0790000 & 0.0896000 & 0.0738 & 0.0708 & 0.0990 & 0.0964\\
500 & 0.0574000 & 0.0604000 & 0.0706000 & 0.0832000 & 0.0600 & 0.0628 & 0.0884 & 0.0918\\
1000 & 0.0518000 & 0.0546000 & 0.0648000 & 0.0754000 & 0.0542 & 0.0556 & 0.0850 & 0.0892\\
\addlinespace[.3em]
\multicolumn{9}{c}{\textit{Early smooth variance reduction, AR errors: $\varphi = .8,\, \vartheta = 0$}}\\
75 & 0.0458000 & 0.0260000 & 0.0512000 & 0.0388000 & 0.0474 & 0.0256 & 0.0726 & 0.0448\\
100 & 0.0526000 & 0.0354000 & 0.0586000 & 0.0496000 & 0.0412 & 0.0272 & 0.0736 & 0.0478\\
150 & 0.0448000 & 0.0364073 & 0.0562000 & 0.0516103 & 0.0438 & 0.0328 & 0.0792 & 0.0610\\
250 & 0.0502000 & 0.0468000 & 0.0624000 & 0.0696000 & 0.0500 & 0.0426 & 0.0828 & 0.0708\\
500 & 0.0502000 & 0.0476000 & 0.0674000 & 0.0720000 & 0.0516 & 0.0474 & 0.0888 & 0.0832\\
1000 & 0.0570114 & 0.0530106 & 0.0708142 & 0.0808162 & 0.0498 & 0.0438 & 0.0826 & 0.0828\\
\addlinespace[.3em]
\multicolumn{9}{c}{\textit{Late smooth variance increase, AR errors: $\varphi = -.8,\, \vartheta = 0$}}\\
75 & 0.0860000 & 0.0880000 & 0.1476000 & 0.1196000 & 0.0556 & 0.0698 & 0.0608 & 0.0732\\
100 & 0.0824000 & 0.0890000 & 0.1516000 & 0.1286000 & 0.0576 & 0.0680 & 0.0644 & 0.0764\\
150 & 0.0756000 & 0.0770000 & 0.1532000 & 0.1234000 & 0.0558 & 0.0674 & 0.0648 & 0.0784\\
250 & 0.0674000 & 0.0716000 & 0.1506000 & 0.1328000 & 0.0474 & 0.0552 & 0.0638 & 0.0708\\
500 & 0.0584000 & 0.0660000 & 0.1592000 & 0.1284000 & 0.0532 & 0.0576 & 0.0774 & 0.0818\\
1000 & 0.0542000 & 0.0554000 & 0.1474000 & 0.1250000 & 0.0514 & 0.0540 & 0.0814 & 0.0894\\
\addlinespace[.3em]
\multicolumn{9}{c}{\textit{Late smooth variance increase, AR errors: $\varphi = .8,\, \vartheta = 0$}}\\
75 & 0.0884000 & 0.0732000 & 0.1580000 & 0.1318000 & 0.0444 & 0.0764 & 0.0556 & 0.0950\\
100 & 0.0770000 & 0.0670000 & 0.1610000 & 0.1186000 & 0.0444 & 0.0672 & 0.0610 & 0.0874\\
150 & 0.0694000 & 0.0614000 & 0.1540000 & 0.1226000 & 0.0468 & 0.0580 & 0.0656 & 0.0854\\
250 & 0.0624000 & 0.0528000 & 0.1566000 & 0.1168000 & 0.0492 & 0.0488 & 0.0750 & 0.0784\\
500 & 0.0616000 & 0.0514000 & 0.1566000 & 0.1222000 & 0.0506 & 0.0492 & 0.0784 & 0.0842\\
1000 & 0.0506000 & 0.0496000 & 0.1518000 & 0.1246000 & 0.0484 & 0.0472 & 0.0820 & 0.0806\\
\bottomrule
\end{tabular}
}
\begin{minipage}{\textwidth}
    \vspace{.25cm}
    \scriptsize\textit{Notes:} DGP \eqref{eq:ardgp_BALURT} with $c=0$ and $\sigma_{T,\,t}$ as defined in \eqref{eq:svt_BALURT}, with $\kappa=.2$, $s_2^2 = .25$ for early negative shifts and $\kappa=.8$, $s_2^2=4$ for late positive shifts. The data are adjusted for a constant ($\vz_t = 1$) or a linear time trend ($\vz_t = (1,t)'$) using the FD method of \textcite{SchmidtPhillips1992}. All lag orders are computed using the RSMAIC. $B=499$ wild bootstrap replications with $q=0$. $5000$ Monte Carlo replications.
    \end{minipage}
\end{table}

\begin{table}[ht]
\setlength{\tabcolsep}{16pt}
\renewcommand{\arraystretch}{1.1}
\centering
\caption{Size of the SWB activation knot tests under heteroskedastic AR errors}
\label{tab:tab_TA1_BALURT}
\vspace{.3cm}
\resizebox{\textwidth}{!}{
	
\begin{tabular}{l *{8}{S} }
\toprule
\toprule
\multicolumn{1}{c}{} & \multicolumn{4}{c}{$\vz_t=1$} & \multicolumn{4}{c}{$\vz_t=(1,t)'$} \\
\cmidrule(l{3pt}r{3pt}){2-5} \cmidrule(l{3pt}r{3pt}){6-9}
$T$ & $\tau^*$ & $\Breve{\tau}^*$ & $\tau$ & $\Breve{\tau}$ & $\tau^*$ & $\Breve{\tau}^*$ & $\tau$ & $\Breve{\tau}$\\
\midrule
\addlinespace[.3em]
\multicolumn{9}{c}{\textit{Early smooth variance reduction, AR errors: $\varphi = -.8,\, \vartheta = 0$}}\\
75 & 0.0494000 & 0.0522000 & 0.0900000 & 0.0892000 & 0.0512 & 0.0606 & 0.1548 & 0.1346\\
100 & 0.0570000 & 0.0560000 & 0.0914000 & 0.0880000 & 0.0470 & 0.0552 & 0.1264 & 0.1232\\
150 & 0.0518000 & 0.0550000 & 0.0814000 & 0.0834000 & 0.0430 & 0.0524 & 0.1096 & 0.1090\\
250 & 0.0562000 & 0.0566000 & 0.0790000 & 0.0896000 & 0.0474 & 0.0480 & 0.0990 & 0.0964\\
500 & 0.0504000 & 0.0530000 & 0.0706000 & 0.0832000 & 0.0480 & 0.0496 & 0.0884 & 0.0918\\
1000 & 0.0494000 & 0.0492000 & 0.0648000 & 0.0754000 & 0.0454 & 0.0498 & 0.0850 & 0.0892\\
\addlinespace[.3em]
\multicolumn{9}{c}{\textit{Early smooth variance reduction, AR errors: $\varphi = .8,\, \vartheta = 0$}}\\
75 & 0.0456000 & 0.0514000 & 0.0512000 & 0.0388000 & 0.0532 & 0.0548 & 0.0726 & 0.0448\\
100 & 0.0536000 & 0.0562000 & 0.0586000 & 0.0496000 & 0.0532 & 0.0544 & 0.0736 & 0.0478\\
150 & 0.0522000 & 0.0528106 & 0.0562000 & 0.0516103 & 0.0546 & 0.0566 & 0.0792 & 0.0610\\
250 & 0.0542000 & 0.0614000 & 0.0624000 & 0.0696000 & 0.0532 & 0.0586 & 0.0828 & 0.0708\\
500 & 0.0552000 & 0.0560000 & 0.0674000 & 0.0720000 & 0.0534 & 0.0566 & 0.0888 & 0.0832\\
1000 & 0.0576115 & 0.0568114 & 0.0708142 & 0.0808162 & 0.0502 & 0.0488 & 0.0826 & 0.0828\\
\addlinespace[.3em]
\multicolumn{9}{c}{\textit{Late smooth variance increase, AR errors: $\varphi = -.8,\, \vartheta = 0$}}\\
75 & 0.0572000 & 0.0502000 & 0.1476000 & 0.1196000 & 0.0260 & 0.0336 & 0.0608 & 0.0732\\
100 & 0.0612000 & 0.0524000 & 0.1516000 & 0.1286000 & 0.0310 & 0.0378 & 0.0644 & 0.0764\\
150 & 0.0600000 & 0.0540000 & 0.1532000 & 0.1234000 & 0.0420 & 0.0450 & 0.0648 & 0.0784\\
250 & 0.0590000 & 0.0540000 & 0.1506000 & 0.1328000 & 0.0360 & 0.0404 & 0.0638 & 0.0708\\
500 & 0.0534000 & 0.0532000 & 0.1592000 & 0.1284000 & 0.0490 & 0.0490 & 0.0774 & 0.0818\\
1000 & 0.0522000 & 0.0498000 & 0.1474000 & 0.1250000 & 0.0478 & 0.0474 & 0.0814 & 0.0894\\
\addlinespace[.3em]
\multicolumn{9}{c}{\textit{Late smooth variance increase, AR errors: $\varphi = .8,\, \vartheta = 0$}}\\
75 & 0.0666000 & 0.0588000 & 0.1580000 & 0.1318000 & 0.0560 & 0.0610 & 0.0556 & 0.0950\\
100 & 0.0582000 & 0.0540000 & 0.1610000 & 0.1186000 & 0.0560 & 0.0558 & 0.0610 & 0.0874\\
150 & 0.0554000 & 0.0564000 & 0.1540000 & 0.1226000 & 0.0530 & 0.0582 & 0.0656 & 0.0854\\
250 & 0.0580000 & 0.0522000 & 0.1566000 & 0.1168000 & 0.0526 & 0.0516 & 0.0750 & 0.0784\\
500 & 0.0594000 & 0.0552000 & 0.1566000 & 0.1222000 & 0.0526 & 0.0498 & 0.0784 & 0.0842\\
1000 & 0.0480000 & 0.0514000 & 0.1518000 & 0.1246000 & 0.0520 & 0.0502 & 0.0820 & 0.0806\\
\bottomrule
\end{tabular}
}
\begin{minipage}{\textwidth}
    \vspace{.25cm}
    \scriptsize\textit{Notes:} DGP \eqref{eq:ardgp_BALURT} with $c=0$ and $\sigma_{T,\,t}$ as defined in \eqref{eq:svt_BALURT}, with $\kappa=.2$, $s_2^2 = .25$ for early negative shifts and $\kappa=.8$, $s_2^2=4$ for late positive shifts. The data are adjusted for a constant ($\vz_t = 1$) or a linear time trend ($\vz_t = (1,t)'$) using the FD method of \textcite{SchmidtPhillips1992}. All lag orders are computed using the RSMAIC. $B=499$ sieve wild bootstrap replications. $5000$ Monte Carlo replications.
    \end{minipage}
\end{table}

\begin{table}[t]
\setlength{\tabcolsep}{16pt}
\renewcommand{\arraystretch}{1.1}
\centering
\caption{Local power of the SWB tests under heteroskedastic AR errors}
\label{tab:tab_TA3_BALURT}
\vspace{.3cm}
\resizebox{\textwidth}{!}{
	
\begin{tabular}{l *{8}{S} }
\toprule
\toprule
\multicolumn{1}{c}{} & \multicolumn{4}{c}{$\vz_t=1$} & \multicolumn{4}{c}{$\vz_t=(1,t)'$} \\
\cmidrule(l{3pt}r{3pt}){2-5} \cmidrule(l{3pt}r{3pt}){6-9}
$T$ & $\tau^*$ & $\Breve{\tau}^*$ & $\tau$ & $\Breve{\tau}$ & $\tau^*$ & $\Breve{\tau}^*$ & $\tau$ & $\Breve{\tau}$\\
\midrule
\addlinespace[.3em]
\multicolumn{9}{c}{\textit{Early smooth variance reduction, AR errors: $\varphi = -.8,\, \vartheta = 0$}}\\
75 & 0.1000 & 0.1128 & 0.1052 & 0.1126 & 0.1284 & 0.1882000 & 0.1248 & 0.1760\\
100 & 0.0992 & 0.1088 & 0.0956 & 0.1000 & 0.1534 & 0.1888000 & 0.1744 & 0.1834\\
150 & 0.0990 & 0.1110 & 0.0986 & 0.1020 & 0.1482 & 0.1862000 & 0.1756 & 0.1916\\
250 & 0.1078 & 0.1164 & 0.1022 & 0.1044 & 0.1652 & 0.1962000 & 0.1744 & 0.2068\\
500 & 0.1256 & 0.1298 & 0.1248 & 0.1262 & 0.1924 & 0.2140000 & 0.2052 & 0.2260\\
1000 & 0.1468 & 0.1538 & 0.1522 & 0.1548 & 0.2156 & 0.2348000 & 0.2268 & 0.2424\\
\addlinespace[.3em]
\multicolumn{9}{c}{\textit{Early smooth variance reduction, AR errors: $\varphi = .8,\, \vartheta = 0$}}\\
75 & 0.1134 & 0.1316 & 0.1068 & 0.1276 & 0.1326 & 0.1416000 & 0.1126 & 0.1464\\
100 & 0.1000 & 0.1142 & 0.0864 & 0.1024 & 0.1346 & 0.1426000 & 0.1350 & 0.1510\\
150 & 0.1170 & 0.1268 & 0.1136 & 0.1248 & 0.1486 & 0.1646000 & 0.1426 & 0.1640\\
250 & 0.1188 & 0.1256 & 0.1112 & 0.1134 & 0.1684 & 0.1782000 & 0.1580 & 0.1706\\
500 & 0.1298 & 0.1350 & 0.1256 & 0.1260 & 0.1874 & 0.2058000 & 0.1764 & 0.1964\\
1000 & 0.1578 & 0.1694 & 0.1372 & 0.1524 & 0.2174 & 0.2378000 & 0.2202 & 0.2446\\
\addlinespace[.3em]
\multicolumn{9}{c}{\textit{Late smooth variance increase, AR errors: $\varphi = -.8,\, \vartheta = 0$}}\\
75 & 0.1396 & 0.2002 & 0.1408 & 0.2244 & 0.1226 & 0.1834367 & 0.1860 & 0.2278\\
100 & 0.1376 & 0.2026 & 0.1268 & 0.2172 & 0.1400 & 0.1886000 & 0.2054 & 0.2454\\
150 & 0.1316 & 0.2068 & 0.1164 & 0.2066 & 0.1670 & 0.2098000 & 0.2090 & 0.2418\\
250 & 0.1340 & 0.2282 & 0.1140 & 0.2300 & 0.1912 & 0.2390000 & 0.2508 & 0.2982\\
500 & 0.1500 & 0.2632 & 0.1444 & 0.2694 & 0.2354 & 0.2778000 & 0.2550 & 0.2858\\
1000 & 0.1560 & 0.2652 & 0.1530 & 0.2730 & 0.2616 & 0.3004000 & 0.2754 & 0.3172\\
\addlinespace[.3em]
\multicolumn{9}{c}{\textit{Late smooth variance increase, AR errors: $\varphi = .8,\, \vartheta = 0$}}\\
75 & 0.0722 & 0.1258 & 0.0522 & 0.0918 & 0.1292 & 0.1354000 & 0.1170 & 0.0888\\
100 & 0.0932 & 0.1514 & 0.0750 & 0.1206 & 0.1410 & 0.1510000 & 0.1332 & 0.1208\\
150 & 0.1046 & 0.1764 & 0.0906 & 0.1594 & 0.1640 & 0.1810000 & 0.1570 & 0.1586\\
250 & 0.1060 & 0.1928 & 0.0920 & 0.1780 & 0.2010 & 0.2154000 & 0.1966 & 0.2146\\
500 & 0.1226 & 0.2286 & 0.1076 & 0.2136 & 0.2322 & 0.2502000 & 0.2284 & 0.2594\\
1000 & 0.1524 & 0.2652 & 0.1646 & 0.2618 & 0.2662 & 0.2938000 & 0.2724 & 0.3068\\
\bottomrule
\end{tabular}
}
\begin{minipage}{\textwidth}
    \vspace{.25cm}
    \scriptsize\textit{Notes:} DGP \eqref{eq:ardgp_BALURT} with $c=-7$ for $\vz_t = 1$ and $c=-13.5$ for $\vz_t = (1,\,t)'$. $\sigma_{T,\,t}$ as defined in \eqref{eq:svt_BALURT}, with $\kappa=.2$, $s_2^2 = .25$ for early negative shifts and $\kappa=.8$, $s_2^2=4$ for late positive shifts. The data are adjusted for a constant ($\vz_t = 1$) or a linear time trend ($\vz_t = (1,t)'$) using the FD method of \textcite{SchmidtPhillips1992}. Estimates for $\tau$ and $\Breve{\tau}$ are size-adjusted at 5\%. All lag orders are computed using the RSMAIC. $B=499$ sieve wild bootstrap replications. $5000$ Monte Carlo replications.
    \end{minipage}
\end{table}

\begin{table}[t]
\setlength{\tabcolsep}{16pt}
\renewcommand{\arraystretch}{1.1}
\centering
\caption{Local power of the SWB tests under heteroskedastic MA errors}
\label{tab:tab_TA4_BALURT}
\vspace{.3cm}
\resizebox{\textwidth}{!}{
	
\begin{tabular}{l *{8}{S} }
\toprule
\toprule
\multicolumn{1}{c}{} & \multicolumn{4}{c}{$\vz_t=1$} & \multicolumn{4}{c}{$\vz_t=(1,t)'$} \\
\cmidrule(l{3pt}r{3pt}){2-5} \cmidrule(l{3pt}r{3pt}){6-9}
$T$ & $\tau^*$ & $\Breve{\tau}^*$ & $\tau$ & $\Breve{\tau}$ & $\tau^*$ & $\Breve{\tau}^*$ & $\tau$ & $\Breve{\tau}$\\
\midrule
\addlinespace[.3em]
\multicolumn{9}{c}{\textit{Early smooth variance reduction, MA errors: $\varphi = 0,\, \vartheta = -.8$}}\\
75 & 0.2054 & 0.2792559 & 0.0776 & 0.0972194 & 0.5000 & 0.6372000 & 0.1130 & 0.1334\\
100 & 0.1654 & 0.2222000 & 0.0702 & 0.0886000 & 0.4106 & 0.5405081 & 0.1284 & 0.1646\\
150 & 0.1348 & 0.1768000 & 0.0800 & 0.1016000 & 0.2794 & 0.3974000 & 0.1326 & 0.1824\\
250 & 0.1370 & 0.1584000 & 0.0854 & 0.1022000 & 0.2020 & 0.2972000 & 0.1456 & 0.1842\\
500 & 0.1400 & 0.1542000 & 0.1082 & 0.1210000 & 0.2008 & 0.2504000 & 0.1580 & 0.1826\\
1000 & 0.1658 & 0.1738000 & 0.1148 & 0.1198000 & 0.2218 & 0.2570000 & 0.1660 & 0.2020\\
\addlinespace[.3em]
\multicolumn{9}{c}{\textit{Early smooth variance reduction, MA errors: $\varphi = 0,\, \vartheta = .8$}}\\
75 & 0.1060 & 0.1142000 & 0.0922 & 0.1038000 & 0.1282 & 0.1570000 & 0.1240 & 0.1562\\
100 & 0.1074 & 0.1198000 & 0.1100 & 0.1142000 & 0.1270 & 0.1508000 & 0.1234 & 0.1580\\
150 & 0.1070 & 0.1208242 & 0.1034 & 0.1100220 & 0.1314 & 0.1576000 & 0.1530 & 0.1752\\
250 & 0.1104 & 0.1258000 & 0.1074 & 0.1098000 & 0.1564 & 0.1836000 & 0.1618 & 0.1970\\
500 & 0.1242 & 0.1322000 & 0.1202 & 0.1266000 & 0.1822 & 0.2002000 & 0.1844 & 0.2004\\
1000 & 0.1574 & 0.1658000 & 0.1482 & 0.1496000 & 0.2060 & 0.2374000 & 0.2232 & 0.2464\\
\addlinespace[.3em]
\multicolumn{9}{c}{\textit{Late smooth variance increase, MA errors: $\varphi = 0,\, \vartheta = -.8$}}\\
75 & 0.4442 & 0.4728000 & 0.1860 & 0.1932000 & 0.5004 & 0.6152000 & 0.1886 & 0.2296\\
100 & 0.3748 & 0.4044000 & 0.1758 & 0.1844000 & 0.4124 & 0.5262000 & 0.1964 & 0.2440\\
150 & 0.3136 & 0.3624725 & 0.1908 & 0.2008000 & 0.2752 & 0.3828000 & 0.1912 & 0.2326\\
250 & 0.3006 & 0.3434000 & 0.1920 & 0.2226000 & 0.2044 & 0.3040000 & 0.1766 & 0.2412\\
500 & 0.3024 & 0.3242000 & 0.2344 & 0.2440000 & 0.1954 & 0.2754000 & 0.1958 & 0.2388\\
1000 & 0.3090 & 0.3280000 & 0.2282 & 0.2376000 & 0.2178 & 0.2870000 & 0.1996 & 0.2558\\
\addlinespace[.3em]
\multicolumn{9}{c}{\textit{Late smooth variance increase, MA errors: $\varphi = 0,\, \vartheta = .8$}}\\
75 & 0.0872 & 0.1566000 & 0.0636 & 0.1340000 & 0.1366 & 0.1606000 & 0.1634 & 0.1446\\
100 & 0.0890 & 0.1646000 & 0.0668 & 0.1598000 & 0.1372 & 0.1658000 & 0.1560 & 0.1682\\
150 & 0.1018 & 0.1882000 & 0.0834 & 0.1806000 & 0.1512 & 0.1796000 & 0.1690 & 0.1688\\
250 & 0.1104 & 0.2060000 & 0.0852 & 0.1822000 & 0.1838 & 0.2058000 & 0.1958 & 0.2216\\
500 & 0.1290 & 0.2430000 & 0.1090 & 0.2444000 & 0.2186 & 0.2456000 & 0.2132 & 0.2482\\
1000 & 0.1516 & 0.2710000 & 0.1412 & 0.2468000 & 0.2382 & 0.2784000 & 0.2392 & 0.2718\\
\bottomrule
\end{tabular}
}
\begin{minipage}{\textwidth}
    \vspace{.25cm}
    \scriptsize\textit{Notes:} DGP \eqref{eq:ardgp_BALURT} with $c=-7$ for $\vz_t = 1$ and $c=-13.5$ for $\vz_t = (1,\,t)'$. $\sigma_{T,\,t}$ as defined in \eqref{eq:svt_BALURT}, with $\kappa=.2$, $s_2^2 = .25$ for early negative shifts and $\kappa=.8$, $s_2^2=4$ for late positive shifts. The data are adjusted for a constant ($\vz_t = 1$) or a linear time trend ($\vz_t = (1,t)'$) using the FD method of \textcite{SchmidtPhillips1992}. Estimates for $\tau$ and $\Breve{\tau}$ are size-adjusted at 5\%. All lag orders are computed using the RSMAIC, setting $q=p$. $B=499$ sieve wild bootstrap replications. $5000$ Monte Carlo replications.
    \end{minipage}
\end{table}
\clearpage
\printbibliography
\end{document}